\documentclass[letterpaper,11pt]{article}

\pdfoutput=1
\usepackage{jheppub}
\usepackage{graphicx,epstopdf,amsmath,amsfonts,amssymb,appendix} % I cant compile with ,bbold
\usepackage{color, xcolor, slashed, setspace, multirow, longtable, braket}
\usepackage{epsfig,mathrsfs,latexsym,color,url,etoolbox, float}
\usepackage{booktabs, subcaption}
\definecolor{nicered}{rgb}{0.7,0.1,0.1}
\definecolor{nicegreen}{rgb}{0.1,0.5,0.1}
\definecolor{CarnationPink}{rgb}{1.0, 0.65, 0.79}
\DeclareMathAlphabet{\mathbbold}{U}{bbold}{m}{n}    

\usepackage{comment}
\usepackage{hyperref}

\usepackage[english]{babel} % hyphenation and typographical rules for a language
\usepackage[utf8]{inputenc} % input encoding
\usepackage[T1]{fontenc} % fonts encoding

\DeclareMathAlphabet{\mathpzc}{OT1}{pzc}{m}{it}

\newcommand{\GeV}{{\rm GeV}}
\newcommand{\TeV}{{\rm TeV}}
\newcommand{\cevns}{CE$\nu$NS }

\definecolor{corurl}{RGB}{0,139,139}
\definecolor{corlinks}{RGB}{39,64,139}
\definecolor{corcite}{RGB}{139,0,0}
\definecolor{kjkblue}{rgb}{0.39, 0.589, 0.6914}
\definecolor{yfpgcol}{RGB}{077, 041, 131}

\def\Fermilab{Theoretical Physics Department, Fermilab, P.O. Box 500, Batavia, IL 60510, USA}
\def\Northwestern{Department of Physics \& Astronomy, Northwestern University, Evanston, IL 60208, USA}
\def\COFI{Colegio de Física Fundamental e Interdisciplinaria de las Américas (COFI), 254 Norzagaray street, San Juan, Puerto Rico 00901}
\makeindex

\graphicspath{{Figures/}}
\begin{document}

\title{LEvEL: Low-Energy Neutrino Experiment at the LHC}

\author[1]{Kevin J. Kelly,}
\author[1]{Pedro A.N. Machado,}
\author[1]{Alberto Marchionni,}
\author[1,2,3]{Yuber F. Perez-Gonzalez}
\affiliation[1]{\Fermilab}
\affiliation[2]{\Northwestern}
\affiliation[3]{\COFI}
\emailAdd{kkelly12@fnal.gov}
\emailAdd{pmachado@fnal.gov}
\emailAdd{yfperezg@northwestern.edu}
\emailAdd{alberto@fnal.gov}

\date\today
\abstract{
We propose the operation of LEvEL, the Low-Energy Neutrino Experiment at the LHC, a neutrino detector near the Large Hadron Collider Beam Dump.
Such a detector is capable of exploring an intense, low-energy neutrino flux and can measure neutrino cross sections that have previously never been observed.
These cross sections can inform other future neutrino experiments, such as those aiming to observe neutrinos from supernovae, allowing such measurements to accomplish their fundamental physics goals.
We perform detailed simulations to determine neutrino production at the LHC beam dump, as well as neutron and muon backgrounds.
Measurements at a few to ten percent precision of  neutrino-argon charged current and neutrino-nucleus coherent scattering cross sections are attainable with 100~ton-year  and 1~ton-year exposures at LEvEL, respectively, concurrent with the operation of the High Luminosity LHC. 
We also estimate signal and backgrounds for an experiment exploiting the forward direction of the LHC beam dump, which could measure neutrinos above 100~GeV.}

\preprint{FERMILAB-PUB-21-046-T, NUHEP-TH/21-01}
\setcounter{tocdepth}{3}

\maketitle

\section{Introduction}

In the wake of the discovery that neutrinos change flavors and therefore have mass, an exciting experimental landscape has developed to better understand their properties. 
These experiments take advantage of a wide variety of neutrino sources; both human-generated like reactor antineutrinos and accelerator beam neutrinos, as well as environmental, such as solar neutrinos, geoneutrinos, and high-energy cosmogenic neutrinos. 
Moreover, a broad set of neutrino detectors, with distinct experimental characteristics are being and will be deployed as part of this effort.
This rich experimental landscape will allow for several novel capabilities in probing neutrino physics. These include (for instance), (1) mitigation of systematic uncertainties with $\nu$PRISM at HK~\cite{Hyper-Kamiokande:2018ofw} and DUNE-PRISM~\cite{vilela_cristovao_2018_2642370}, (2) detailed topological information in liquid argon time projection chambers (LArTPCs)~\cite{Acciarri:2018myr,Foreman:2019dzm,Abratenko:2020sga}, (3) low-energy detection thresholds~\cite{Akimov:2017ade,Akimov:2020pdx,Fernandez-Moroni:2020yyl}, and many more opportunities.

As this field develops, so do its goals. 
Among those goals, a crucial point of current and future large volume neutrino experiments is the study of core collapse supernova~\cite{Janka:2006fh}.
While neutrinos from a supernova have been definitively identified once before by IMB~\cite{Bionta:1987qt,Bratton:1988ww} and Kamiokande-II~\cite{Hirata:1987hu,Hirata:1988ad}, it is expected that, with current \& upcoming experiments~\cite{Abe:2016nxk,Abi:2020evt,An:2015jdp,Abe:2018uyc}, a wealth of knowledge would be acquired if another were to occur in the near future. 
This includes not only knowledge regarding the supernova~\cite{Keil:2002in}, but perhaps also fundamental neutrino properties such as the mass ordering~\cite{Dighe:1999bi, Serpico:2011ir}, stability~\cite{1993APh.....1..377O,Ando:2003ie,deGouvea:2019goq}, self interactions~\cite{Das:2017iuj}, collective oscillations~\cite{Duan:2006an,Fogli:2007bk,Dasgupta:2011jf,Dasgupta:2016dbv,Dasgupta:2018ulw,Chakraborty:2015tfa,Izaguirre:2016gsx,Capozzi:2018clo}, neutrino magnetic moments~\cite{Balantekin:2007xq,deGouvea:2012hg}, among others~\cite{GilBotella:2003sz, EstebanPretel:2007ec, Raffelt:2011nc, deGouvea:2020eqq}.
However, in order to extract many of these interesting features, a precision understanding of low-energy neutrino-nucleus cross sections is required.
In LArTPCs, one of the main detection channels to study supernova is neutrino capture in Argon, $\nu_e\, ^{40}{\rm Ar} \to e^- \, ^{40}{\rm K}^*$~\cite{Bueno:2003ei, Scholberg:2012id}. 
The typical signature of this process, for neutrinos with energies around tens of MeV, is a low energy electron (a short and not necessarily straight) track~\cite{Acciarri:2017sjy} and one or more photons from the de-excitation of $^{40}$K$^*$ (``blips'' in the TPC)~\cite{Castiglioni:2020tsu,Gardiner:2020ulp}. 

Recently, one of the most recent exciting developments in the field of neutrino physics was the first observation of Coherent Elastic Neutrino-Nucleus Scattering (CEvNS)~\cite{Freedman:1973yd} by the COHERENT collaboration~\cite{Akimov:2017ade,Akimov:2020pdx}, in which neutrinos produced at the Spallation Neutron Source (SNS) scatter coherently off the cesium iodide~\cite{Akimov:2017ade} or argon~\cite{Akimov:2020pdx} nuclei in the COHERENT detectors. 
CEvNS offers a new way to observe neutrino interactions and, through precision measurements, will allow for better understanding of neutrinos in the standard model~\cite{Collar:2014lya, Skiba:2020msb, Coloma:2020nhf,Tomalak:2020zfh} and beyond~\cite{Formaggio:2011jt, Pospelov:2011ha, Harnik:2012ni, Anderson:2012pn, Pospelov:2012gm, Dutta:2015nlo, Canas:2017umu, Cui:2017ytb, Ge:2017mcq,Kosmas:2017zbh, Kosmas:2017tsq, Coloma:2017ncl, Farzan:2017xzy, Liao:2017uzy, Esteban:2018ppq, Farzan:2018gtr, Boehm:2018sux, Denton:2018xmq, Cadeddu:2018dux, Billard:2018jnl, AristizabalSierra:2018eqm, Altmannshofer:2018xyo, Abdullah:2018ykz, Bertuzzo:2017tuf,Gonzalez-Garcia:2018dep, Bertuzzo:2018itn, Dutta:2019nbn, Giunti:2019xpr, Dutta:2019eml, Blanco:2019vyp, AristizabalSierra:2019ufd, Dent:2019ueq, Bischer:2019ttk, Canas:2019fjw, Babu:2019mfe, Denton:2020hop, Flores:2020lji, Cadeddu:2020nbr,  Miranda:2020syh}.
While this cross section relatively is large, these scattering events are difficult to observe.
Not only are soft nuclear recoils hard to detect, but also backgrounds, particularly beam related and cosmic-induced neutrons, can mimic the CEvNS signal.

In fact, background rejection (as alluded to in the last two paragraphs) plays a crucial role in experiments that aim to measure low energy neutrino signatures such as CEvNS and neutrino capture on argon.
In this manuscript, we highlight the potential to build on these developments and better our understanding of low-energy neutrino cross sections in a specific environment. 
We propose to leverage the high intensity of the Large Hadron Collider (LHC) to greatly enhance background rejection and probe low energy neutrino signals.
At the LHC, the high-energy $7$ TeV\footnote{We consider the proton energy expected during the high-luminosity LHC (HL-LHC) operation~\cite{Goddard:2017but}.} protons could be aborted twice per day in the LHC beam dump (LHCBD)~\cite{Hoefert:1996nc, Goddard:2017but}. 
The dump of the beam takes in total 86~$\mu$s, which translates into a beam on-off ratio of $2\times10^{-9}$: 100,000 times larger than SNS~\cite{Scholberg:2005qs}.
The dumping of 7~TeV protons results in the production of an enormous number of light hadrons which then decay and produce an intense flux of neutrinos. 
Many of those hadrons will slow down, stop, and decay, leading to a large isotropic flux of low-energy\footnote{Throughout this work, we will often discuss fluxes of both ``low-energy'' and ``high-energy'' neutrinos. In general, we consider neutrinos with $E_\nu \lesssim 10$ GeV to be ``low-energy'' and $E_\nu \gtrsim 10$ GeV to be ``high-energy,'' although this distinction is of course, arbitrary.} neutrinos (below about 250~MeV) from pion-decay-at-rest ($\pi$DAR) and kaon-decay-at-rest ($K$DAR).
Meanwhile, in the forward direction, several mesons decay in flight, resulting in a high-energy flux of neutrinos with energies at the hundreds of GeV.

We propose the operation of one or more LArTPC neutrino detectors near the LHC Beam Dump which we call \textbf{LEvEL}: ``Low-Energy Neutrino Experiment at the LHC''. 
These detectors can be suited for measuring different neutrino scattering processes, and we consider two distinct options in this manuscript. 
One is a liquid argon time-projection chamber similar to SBND~\cite{Machado:2019oxb} or  ProtoDUNE~\cite{Abi:2017aow, Abi:2020mwi}, which would be in an excellent position to observe processes like $\nu_e~^{40}$Ar $\to e^-~^{40}$K$^*$, as well as other neutral- and charged-current quasielastic scattering processes. 
As we will show, a 100~ton detector situated $15$~m from the LHCBD can expect $\mathcal{O}(1-10)$ neutrino capture events \textit{per beam dump}, that is, in 86~$\mu$s.
This is as close as one can get to detecting something similar to two supernovae a day, and would be an excellent experimental setup to prepare for the detection of a core collapse supernova event.
The second option we consider is a $1$~ton detector optimized for observing CEvNS interactions with liquid argon, similar to that used by the COHERENT collaboration. 
This setup would provide a unique high-statistics, low-background measurement of CEvNS.
Although the neutrino flux in the forward direction is quite impressive, we will see that muon backgrounds are prohibitive unless extra mitigation measures are taken.
Because of that we will not focus on this scenario here (see e.g. the  FASER$\nu$~\cite{Abreu:2020ddv} and SND@LHC~\cite{Collaboration:2729015} proposals).
Nevertheless, we will study the dominant backgrounds for both high  and low energy setups in detail below.

This manuscript is organized as follows: in Section~\ref{sec:Setup}, we introduce the LHC Beam Dump and the surrounding environment. 
We explain how we simulate the beam dump to obtain the resulting neutrino flux as well as any particles that can contribute to backgrounds for the searches of interest.
We also show the neutrino fluxes in the forward and off-axis directions.
Section~\ref{sec:Forward} discusses the muon and neutron backgrounds in detail. 
Section~\ref{sec:OffAxis} explores the region perpendicular to the beam direction and determines the low-energy flux from $\pi$DAR and $K$DAR there. 
We demonstrate that this offers a high-signal, low-background region of interest, and propose the various neutrino cross section measurements at this location. 
Finally, in Section~\ref{sec:Conclusions}, we offer some conclusions.

\section{The LHC Beam Dump and Resulting Neutrino Flux}\label{sec:Setup}

In this section we detail our assumptions regarding the LHC Beam Dump and the surrounding environment, including shielding, surrounding soil, and the possibility of a nearby detector. We also provide the resulting neutrino flux at a variety of interesting nearby locations, both for the low-energy and high-energy components of the flux.

\begin{figure}[t]
\centering
\includegraphics[width=0.485\textwidth]{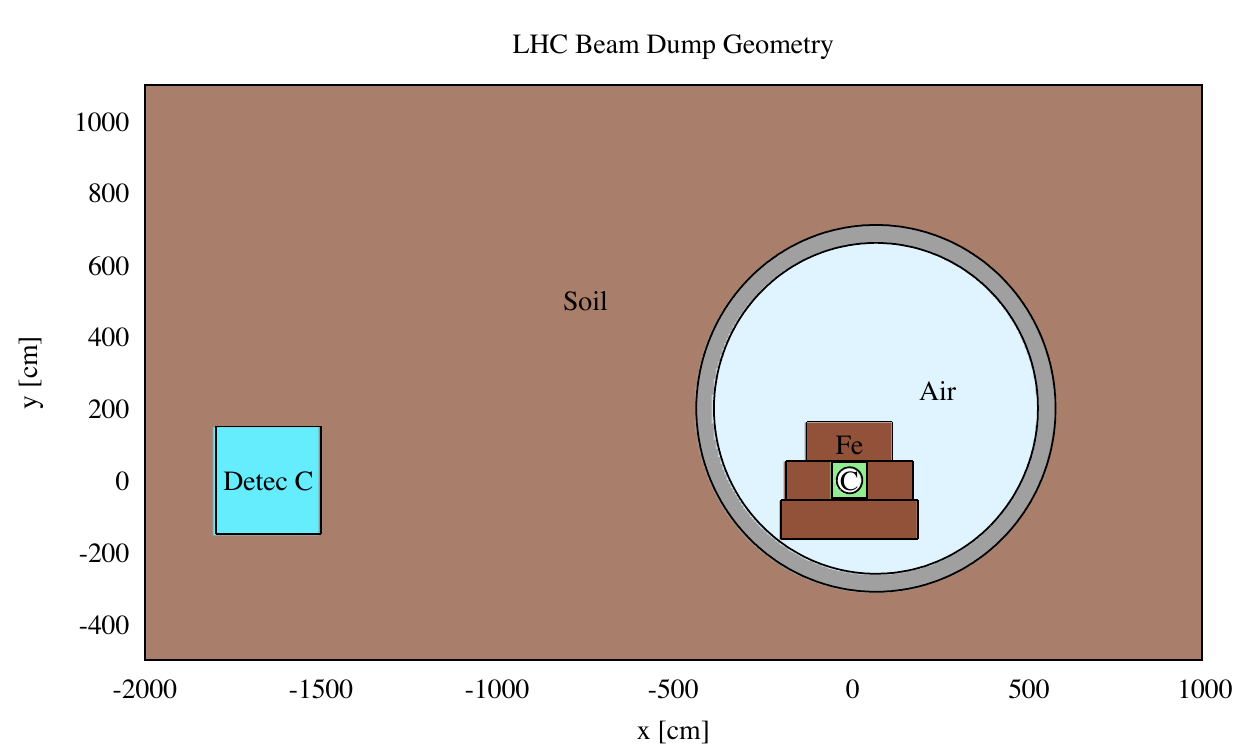}\\
\includegraphics[width=0.485\textwidth]{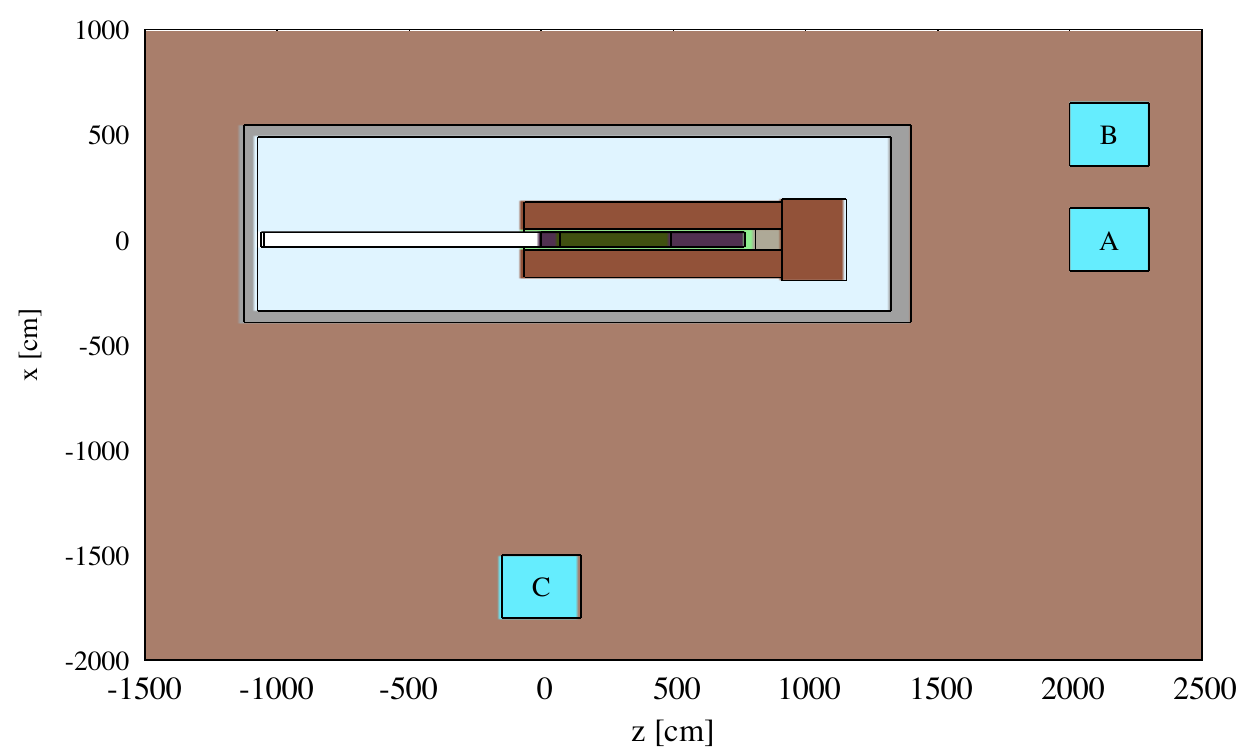}
\includegraphics[width=0.485\textwidth]{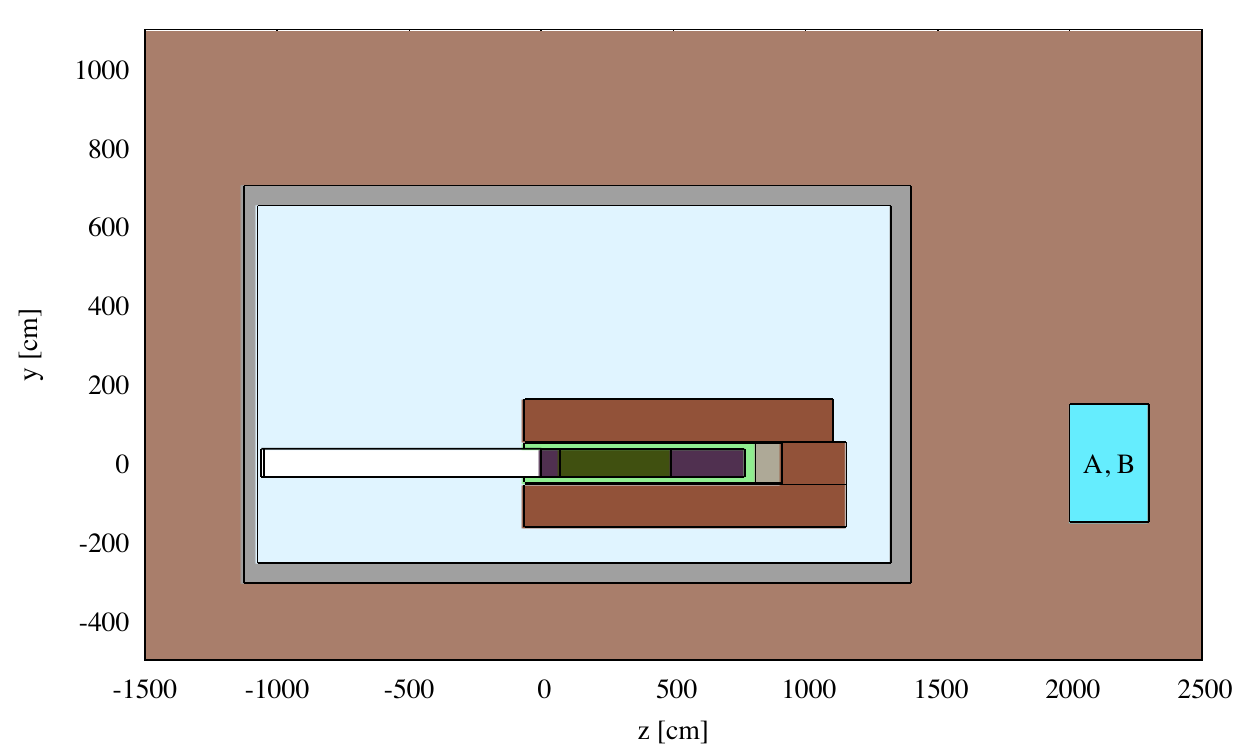}
\caption{Simplified view of the LHC Beam Dump. In the top panel, the circle labelled ``C'' represents the Carbon (graphite) target on which the LHC beam is aborted. Such target is divided into three segments with different densities (green and purple), and it is covered by a steel layer (lighter green). The graphite and its cover is surrounded by iron blocks (dark brown) while in the forward direction there is a Titanium window (light gray). The cavern (which we assume is filled with air) is enclosed by a layer of concrete (gray), with soil (light brown) surrounding it. Three potential detector locations (cyan), labelled ``A'', ``B'', and ``C'' in the different views, are also included. 
}
\label{fig:SchematicBD}
\end{figure}

Details of this setup follow from Ref.~\cite{Goddard:2017but}, and we assume operation is concurrent with the HL-LHC. The LHC Beam Dump (LHCBD) stations, one per circulating beam and located at Insertion Region 6, consist of the target dump external (TDE), a graphite target roughly 70 cm in diameter and 770 cm in length. 
Such a graphite target is surrounded by iron, while the cavern in which the dump station is situated is surrounded by a layer of concrete, all of which is surrounded by soil. 
In what follows, we consider a simplified geometrical description of the LHCBD station.
The center of the face of the graphite target is located at the origin, and the impinging beam is going in the ${+}z$ direction. 
Fig.~\ref{fig:SchematicBD} shows this configuration in three different views. 
As we will discuss in the following, Fig.~\ref{fig:SchematicBD} also shows three potential detector configurations that we consider, at Positions labelled ``A'', ``B'', and ``C''.
%%%%%%%%%%    FIG     %%%%%%%
%%%%%%%%%%%%%%%%%%%%%%%%%%%%%%%%%%%%%%%%%

For each beam abort, $2760$ proton bunches, each with $1.1\times 10^{11}$ protons\footnote{This is the nominal expectation during physics operation of the HL-LHC -- some aborted extractions may have bunches with up to ${\sim}2.2\times 10^{11}$ protons depending on the filling scheme, but we will use this conservative estimate instead.}, are impinged onto the dump, translating into $N_{\rm POT} = 3.0 \times 10^{14}$ protons on target (POT) per dump. 
Each bunch is $0.25$~ns long and separated by $25$ ns. 
Since we will not consider a detector with such fine timing precision, we are then more interested in the overall beam abort time of $86~\mu$s. 
The total stored beam energy that is diverted onto the dump twice a day is $340$ MJ -- this corresponds to an average power (during the beam abort time) of 4.0 TW, or, when averaged over the two-dumps-a-day schedule, $7.9 \times 10^{-3}$ MW~\cite{Metral:2301292}.

In the next sections, we will focus on the  low-energy neutrino flux, which comes predominantly from decays of stopped pions and kaons, as well as some component from even heavier mesons. 
We find it instructive to compare the setup we consider here to other low-energy neutrino sources. 
For a wide variety of low-energy sources, the figure of merit for signal production, in terms of the overall neutrino flux produced, is typically the average beam power.
This can be seen as a combination of both the number of protons on target that a facility can deliver, as well as the energy of said protons. 
The latter is due to the fact that higher-energy protons produce more light mesons per interactions, leading to a high flux of neutrinos (see, e.g., Ref.~\cite{Grant:2015jva}). 
As a direct example, the LHCBD receives a relatively small number of protons-on-target per year:  about $1.2\times10^{17}$~POT/year, considering 200 days of operation yearly. 
For comparison, the J-PARC Materials and Life sciences Facility (MLF) delivers a massive ${\sim}3.8\times 10^{22}$~POT/year~\cite{Ajimura:2017fld}. 
Although at first glance this factor of $10^5$ may seem discouraging, the higher energy of the LHCBD of $7$ TeV compared to the 3 GeV of  MLF brings the figure of merit for signal production closer together. 

The other important metric for these low-energy facilities is related to background rejection.
One of the most straightforward way of rejecting background is by having a very low duty factor, that is, a low ratio of beam-on to beam-off time.
In other words, environmental backgrounds can be rejected by delivering the largest amount of protons on target on the smallest time window possible.
This is where the LHC beam dump excels.
Because the LHCBD aborts the beam relatively quickly, and only dumps its protons twice daily, this corresponds to a very low duty factor.
Therefore, searches for signals only during the beam abort time will have a very high background rejection factor.

\begin{figure}[t]
\centering
\includegraphics[width=0.8\textwidth]{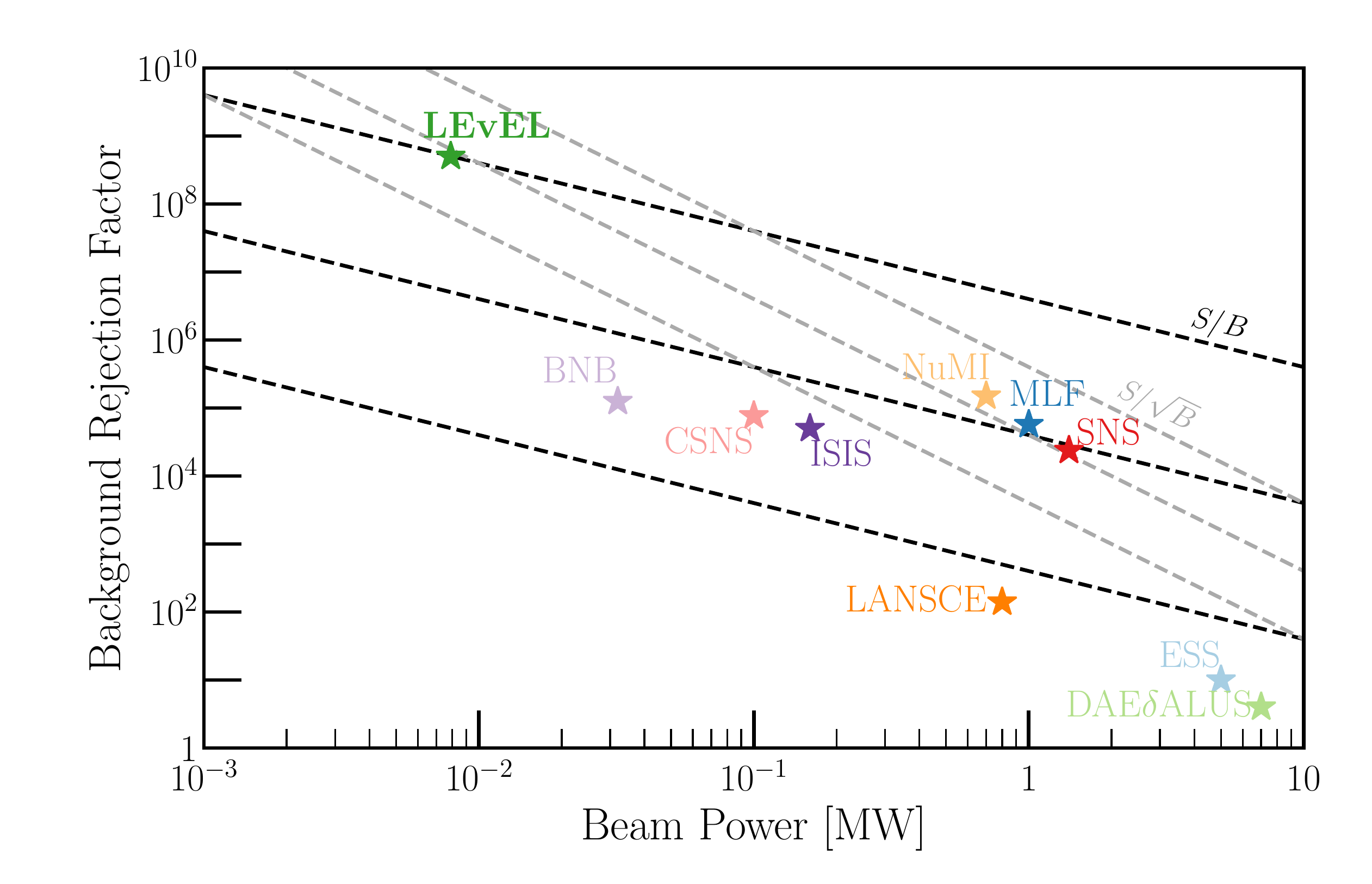}
\caption{Average power vs.~background rejection factor (defined as the ratio between beam-off to beam-on time) for a variety of facilities. 
Our proposal, labelled ``\textbf{LEvEL}'' is represented by the green star. 
Other facilities shown are the Booster Neutrino Beam (faint purple, US)~\cite{AguilarArevalo:2008yp}, China Spallation Neutron Source (pink, China)~\cite{2009NIMPA.600...10W}, ISIS Neutron and Muon Source (purple, UK)~\cite{Burman:1996gt}, Los Alamos Neutron Science Center (orange, US)~\cite{Athanassopoulos:1996ds}, J-PARC Materials and Life sciences Facility (blue, Japan)~\cite{Ajimura:2017fld}, the Neutrinos from the Main Injector beam (NuMI, faint orange, US)~\cite{NuMI}, the Spallation Neutron Source (red, US)~\cite{Scholberg:2005qs}, the European Spallation Source (light blue, Sweden)~\cite{Baussan:2012cw}, and the proposed DAE$\delta$ALUS source (light green, proposed)~\cite{Alonso:2010fs}. 
Dashed black lines indicate our simple assumptions for where signal-to-background ratio $S/B$ is constant, and dashed grey lines are where signal-to-square-root-of-background $S/\sqrt{B}$ is constant.\label{fig:DutyFactor}}
\end{figure}
We compare the LHCBD power and background rejection capability, defined as the inverse of the duty factor, with other stopped pion facilities in Fig.~\ref{fig:DutyFactor}.
In performing this comparison we are focusing on the sources themselves, regardless of the detector technology that would allow to leverage the source advantages and mitigate possible drawbacks. This is a simpler comparison than pairing a large number of possible detector technologies to the many available/planned sources.
The setup considered here is represented by the green star labelled ``\textbf{LEvEL}'', with a low beam power, due to the average over the entire two-dumps-a-day schedule, but very high background rejection factor. 
We also show a wide range of existing and proposed stopped pion sources in this plane, ranging from the Booster Neutrino Beam with low power (but high background rejection factor) to the proposed DAE$\delta$ALUS experiment, with very high power, but low background rejection factor. 
Also shown in Fig.~\ref{fig:DutyFactor} are dashed lines of constant signal-to-background ratio ($S/B$, black dashed lines) or of constant signal-to-square-root-of-background ratio ($S/\sqrt{B}$, grey dashed lines). 
Comparing $S/B$ is useful for comparing precision, systematics-limited measurements of phenomena, where $S/\sqrt{B}$ is a useful metric for discovery in low-statistics environments. 
We see that our proposal is among the best for $S/\sqrt{B}$ and better than all of the current/proposed facilities for $S/B$, with these assumptions regarding signal and background scaling.

Motivated by these figures of merit and the potential of the LHC beam dump for low energy neutrino physics, we proceed to more detailed numerical studies.
We use the {\tt FLUKA} MonteCarlo package together with the {\tt FLAIR} graphic interface \cite{Battistoni:2015epi,Bohlen:2014buj,Flair} to simulate LHC Beam Dump facility and derive the neutrino flux when the beam is aborted. 
We assume a specific setup for the beam dump geometry and chemical composition to give a quantitative estimate of the emitted neutrino flux. 
The main component of the absorption block, its TDE is composed of two different cylinders of polycrystalline graphite with densities of $1.73 {\rm\ g/cm^3}$ with $\{0.7{\rm\ m},\ 3.5{\rm\ m}\}$ of length, respectively, and radius of $0.35{\rm\ m}$; in between these two components, there is cylinder of flexible graphite ($\rho= 1.1 {\rm\ g/cm^3}$) having a length of $3.5{\rm\ m}$~\cite{Goddard:2017but}. 
Around the graphite cylinders, there is a stainless steel vessel with a thickness of $12-14{\rm\ mm}$. 
At the downstream limit of the graphite and few meters upstream the dump core, there is a titanium window (10 mm thick) to contain the graphite in the nitrogen environment. It is then surrounded by iron shielding blocks in order to reduce the radiation produced. The shielding blocks are made of about 35 modified iron yokes with a specific geometry. 
The detailed {\tt FLUKA} inputs we use for our simulations were obtained from the CERN engineering department, and corresponds to the LHCBD geometry used during the runs 1-2. 
The final design for the beam dump during the HL-LHC is still under discussion, so the neutrino fluxes and beam-related backgrounds will depend on the definitive layout. 
Nevertheless, let us stress that the specific details of the geometry will not impact considerably the results we present henceforth.

The TDE at cavern UD62, ${\sim}$100 meters underground, is located downstream of the TD62 tunnel (another TDE is in cavern UD68). 
The cavern can be approximated by a cylinder cut because of the floor with an inside radius of $4.5{\rm\ m}$ and an external radius of $5{\rm\ m}$. 
The cavern wall is  made of concrete and its surroundings are Molasse. 
Civil construction that would occur in the event that LEvEL materializes should be optimized to maximize the passive shielding of the experiment.
Finally, the dumped proton flux is assumed to be time and position independent during the abortion time. 
This is different from what actually occurs in dump, as the beam is swept on the front section of the dump, i.e., each bunch does not superimpose in space to the previous one, to avoid overheating of the TDE. 
The fine grained dependence on time and position of the beam dump should not affect our results.

In order to understand the neutrino flux that arises from the LHC Beam Dump, we consider three hypothetical detector locations. 
The first two, labelled ``A'' and ``B'' in Fig.~\ref{fig:SchematicBD}, are assumed to be 20 meters from the front face of the graphite target. Position A is centered in the beam axis, whereas Position B is 5~m off-axis, that is at an off-axis angle of roughly 14$^\circ$. 
For Position C, we assume a closer detector (with a front face only 15~m from the beam dump target), completely orthogonal to the incoming beam direction. 

\begin{figure}[t]
\centering
\includegraphics[width=\textwidth]{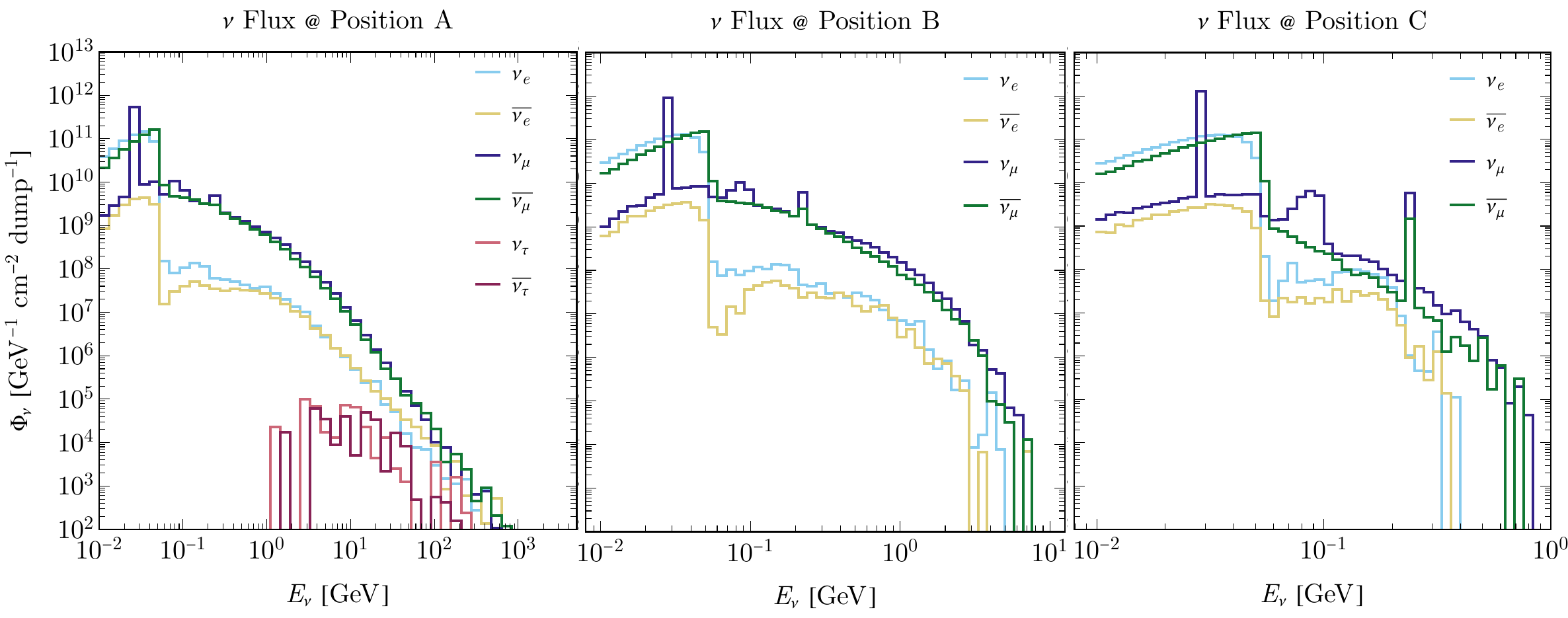}
\caption{Neutrino fluxes per beam dump at Positions A, B, C as labelled in Fig.~\ref{fig:SchematicBD} and explained in the text. Different colors correspond to different flavor of neutrino flux, as well as neutrino and antineutrino components. Note the different scales in the $x$-axis ($E_\nu$) in the three panels -- the flux at Position A extends to significantly higher energy from the decays of very boosted mesons. In contrast, the flux at Position C is described entirely by meson decay-at-rest and muon decays. For reference, we consider an average of two beam dumps per day and 200 days of LHC running per year.
\label{fig:FluxABC}}
\end{figure}

Fig.~\ref{fig:FluxABC} displays the neutrino fluxes, broken down by different flavors of neutrinos, at Position A (left), Position B (center), and Position C (right). 
The MonteCarlo-related uncertainty for these fluxes is at the $2\%$ level.
In all positions we observe that the neutrino flux is dominated by pion ($\pi$DAR), muon ($\mu$DAR), and Kaon ($K$DAR) decays at rest, together with the contribution of muon decay in flight. 
This flux is originated from the interactions of the LHC beam protons with the beam dump material, in such a way that a large quantity of mesons is produced due to the large energy of the incoming flux. 
A large portion of the pions and kaons, produced either directly at the proton-dump interaction or from decays of other mesons, stop in the central graphite cylinder or in the external iron blocks, producing a large flux of neutrinos from decays at rest. 
In Position A we find an additional significant contribution of neutrinos produced by pions and other mesons decaying in flight.
Moreover, given the relatively high energy on the interactions ($\sqrt{s}\sim 100\ \GeV$), we see that the flux in the forward direction at Position A has a very boosted component: neutrinos can reach energies up to ${\cal O}(1\ \TeV)$.  
In this case we have a non-negligible $\nu_\tau$ and $\overline{\nu}_\tau$ flux, produced mainly by $D_s^\pm$ decays. 
In Position B, however, we find a strong reduction of the high energy component of the flux. 
Such reduction is clearly related to the strongly boosted interactions produced by the incoming LHC beam. 
Finally, the neutrino flux in Position C is completely dominated by $\pi$DAR, $K$DAR and low energy muon decays. 
In Appendix~\ref{ap:pflu}, we provide further studies of the high-energy neutrino flux as a function of the off-axis angle and neutrino energy.

%%%%%%%%%%    FIG     %%%%%%%
%%%%%%%%%%%%%%%%%%%%%%%%%%%%%%%%%%%%%%%%%

%%%%%%%%%%%%%%%%%%%%%%%%%%%%%%%%%%%%%%%%%
%%%%%%%%%%%%%%%%%%%%%%%%%%%%%%%%%%%%%%%%%
\section{Muons and Neutrons at LEvEL}\label{sec:Forward}
%%%%%%%%%%%%%%%%%%%%%%%%%%%%%%%%%%%%%%%%%
%%%%%%%%%%%%%%%%%%%%%%%%%%%%%%%%%%%%%%%%%

Although the neutrino fluxes in the left and center panels (Positions A and B) of Fig.~\ref{fig:FluxABC} look very impressive, especially at higher energies, we first need to consider the backgrounds to understand if this setup is indeed promising.
The backgrounds that could affect any measurement at LEvEL will be dependent on the chosen experimental setup, position of the detector, and experimental signature searched for. 
Backgrounds at LEvEL are mostly related to the secondary particles produced after the extraction of the LHC beam. 
Other background sources, such as cosmic rays, can be well-understood and efficiently rejected given the significantly large period of time between dumps. 

\begin{figure}[h]
     \centering
     \begin{subfigure}[b]{0.495\textwidth}
         \centering
         \includegraphics[width=\textwidth]{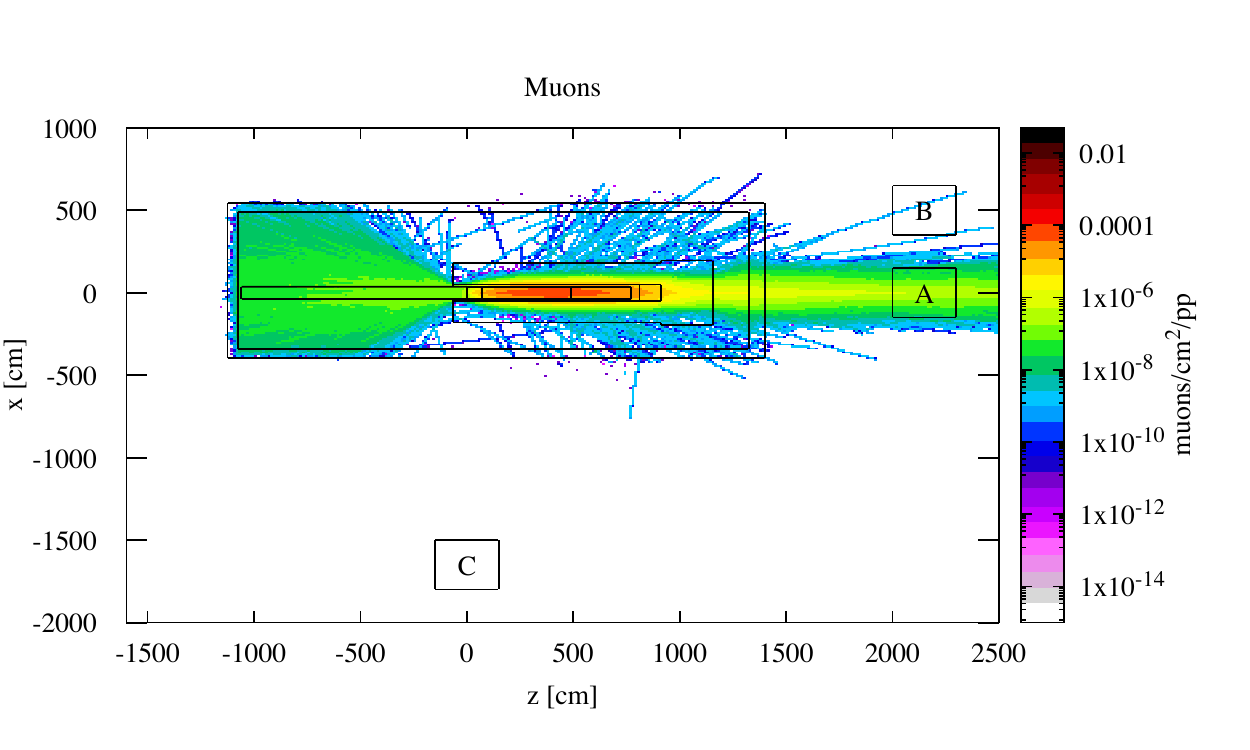}
         \label{fig:muo}
     \end{subfigure}
     \hfill
     \begin{subfigure}[b]{0.495\textwidth}
         \centering
         \includegraphics[width=\textwidth]{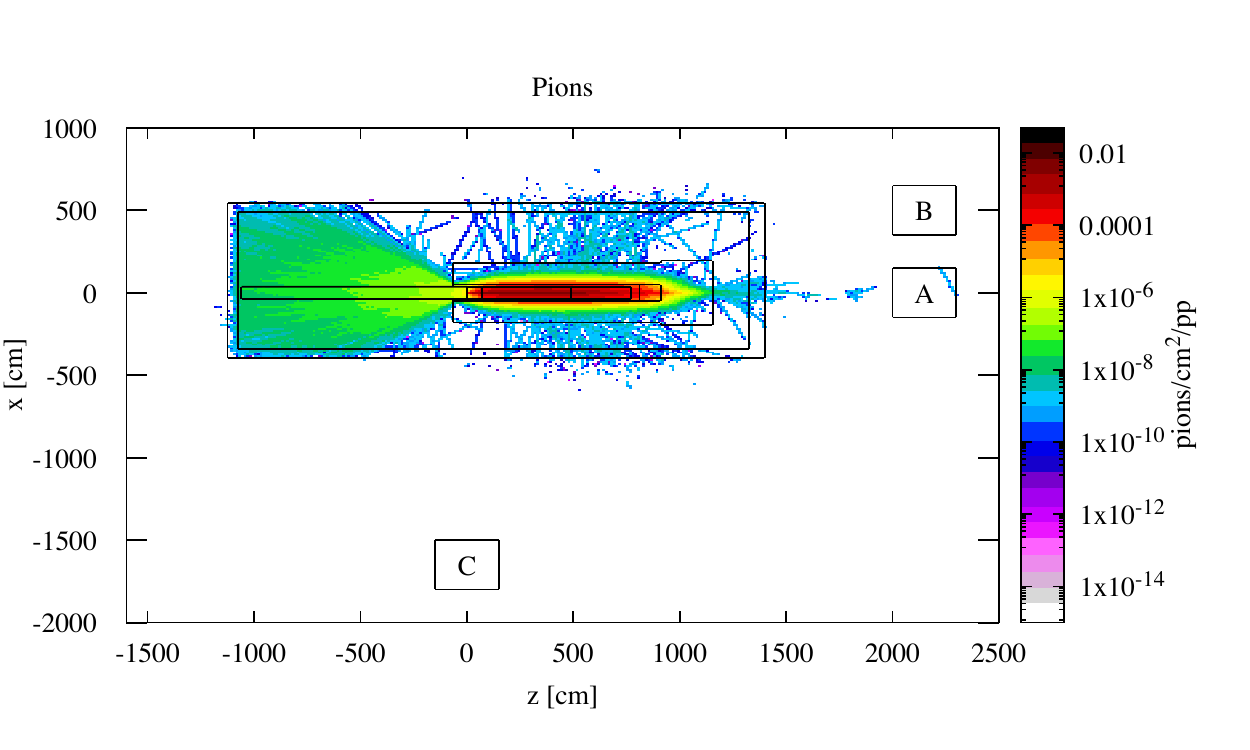}
         \label{fig:pio}
     \end{subfigure}
     \hfill
     \begin{subfigure}[b]{0.495\textwidth}
         \centering
         \includegraphics[width=\textwidth]{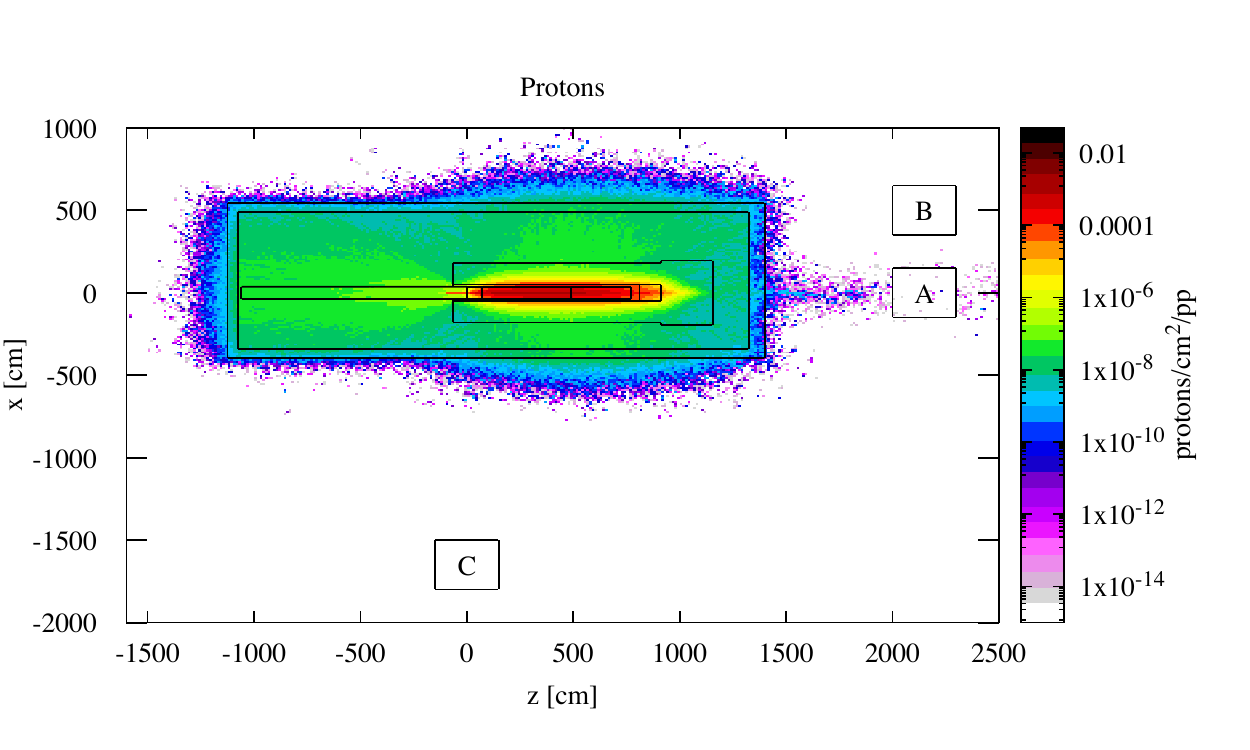}
         \label{fig:pro}
     \end{subfigure}
     \hfill
     \begin{subfigure}[b]{0.495\textwidth}
         \centering
         \includegraphics[width=\textwidth]{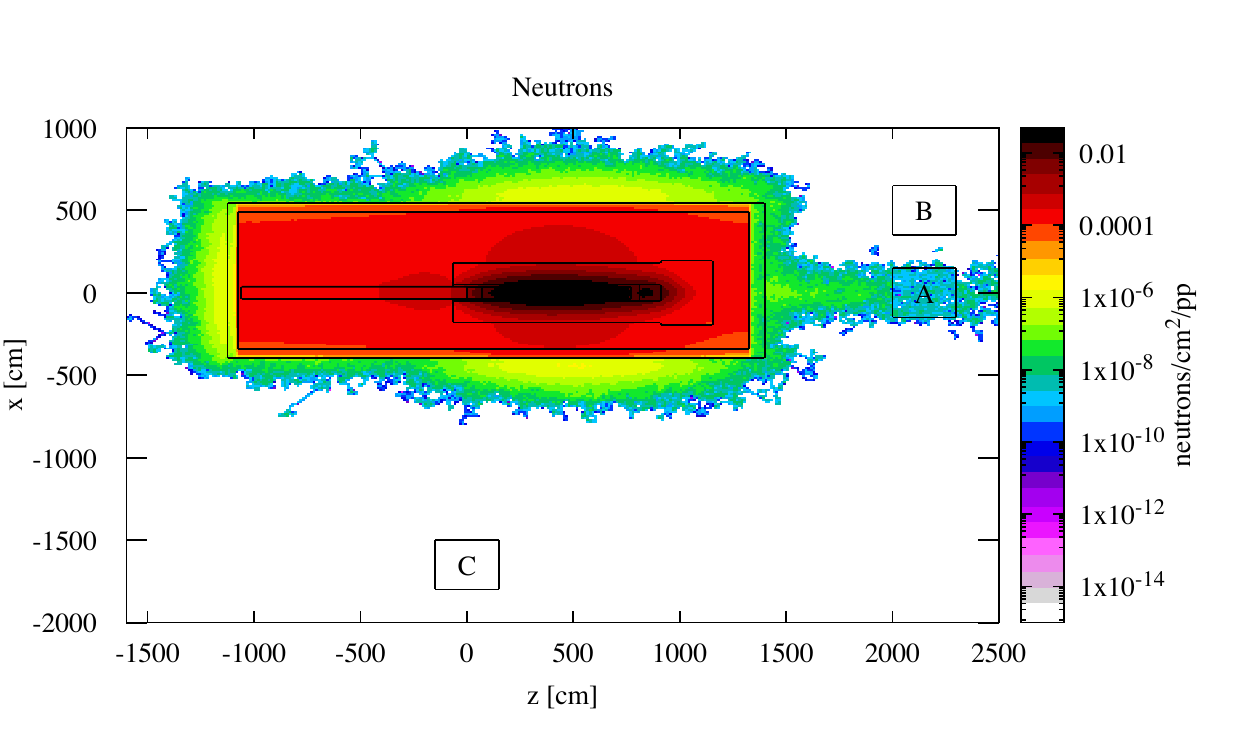}
         \label{fig:neu}
     \end{subfigure}
        \caption{Fluence of background particles: muons (top left), pions (top right), protons (bottom left) and neutrons (bottom right).}
        \label{fig:backf}
\end{figure}

In Fig.~\ref{fig:backf}, we show the fluences\footnote{The fluence in a certain region is defined as the path length of a particle divided by the volume. This serves as a proxy for the flux of a given particle in the area of interest.} for background particles per each dumped proton: muons (top left), pions (top right), protons (bottom left) and neutrons (bottom right).
First we notice that most protons and pions  stop in the concrete walls surrounding the beam dump station.
Muons on the other hand exhibit a high fluence in the forward and backward directions.
The backwards component is low energy, coming mainly from meson decay-at-rest, and thus stops at the concrete wall. 
The forward component, on the other hand, is high energy and has a high penetrating power.
Finally, the neutron fluence is more isotropic, though we can see the presence of forward neutrons which are higher in energy. 
Notice that the overall neutron fluence is higher than the fluence for other particles, which makes neutrons the most challenging background except in the forward direction.

From this, we can conclude that there are two main beam related background categories, which we summarize below:
\begin{enumerate}
  \item\emph{High energy muons.} 
  As a result of the interaction of high energy LHC protons, we can expect the production of very boosted secondary particles. 
  Although the different components of the LHCBD are designed to absorb most of the particles produced, high energy forward muons can escape the facility and reach the LEvEL detector if placed in Position A. Meanwhile, very energetic particles would not affect a detector at Position C  due to its large off-axis angle.
    \item\emph{Low energy neutrons.} 
 Low energy neutrons, subproducts of interactions of secondary particles and nuclear de-excitations, are produced in large quantities.
  Their typical scattering signature in an argon detector is one or more low energy nuclear recoils. This background is thus very relevant for \cevns measurements (see, e.g., the analyses of the COHERENT experiment~\cite{Akimov:2017ade,Akimov:2020pdx}).
  As this flux is nearly isotropic, it constitutes a relevant source of background at all of the proposed positions of LEvEL. 
\end{enumerate}

Regarding the high energy muons, at Position A ($d = 20$ m), the muon flux is found to be  about $10^8 \mu^\pm/{\rm cm}^2$ integrated over each beam dump time, which is an overwhelming rate.
It is impractical to mitigate this large background at Position A  using passive shielding only.
One solution, explored in other experimental proposals such as SHiP~\cite{Anelli:2015pba}, is to place the detector further downstream the beam and employ an active muon shield, which essentially deflects those muons away from the detector. 
Additionally, the proposed/planned FASER$\nu$~\cite{Abreu:2020ddv} and SND@LHC~\cite{Collaboration:2729015} collaborations intend on measuring a flux of high-energy neutrinos produced in the LHC collision point in the forward direction. 
Because this forward direction would require a substantial amount of background mitigation, we defer its focus to future work. 
In Appendix~\ref{ap:pflu}, we provide further considerations on the background in the forward direction, including the on-axis muon and neutron fluxes as a function of the distance from the dump.

The other source of backgrounds we study are neutrons induced after a beam abortion. 
The high energy of the proton beam at the LHC, when dumped, is expected to create a very large number of neutrons due to secondary scatterings.
Although a {\tt FLUKA} simulation can be performed to estimate this background, the large number of secondary particles makes the simulation exceptionally long.
Optimized and realistic neutron mitigation strategies are beyond the scope of this paper.
Thus, we decide to perform a toy simulation to understand how far a detector perpendicular to the beam axis should be located in order to have manageable neutron backgrounds.

The toy simulation consists of a symmetric, cylindrical beam dump cavern. 
For reference, we use $\rho$ for the distance to the central cylinder axis and $z$ for the distance along the beam axis.
The dump is put at the central axis of the cylinder, $\rho=z=0$.
The walls are located at $\rho=5.5$~meters away from the axis~\footnote{Note that the dump station has a diameter of about 10 meters. Nevertheless, the beam dump is not in the center of the cavern. The horizontal distance between the dump and the left (right) wall is of 5.5~m (4.5~m). To be conservative, we use 5.5~m radius in our toy simulation.}.
Surrounding the walls, there is borated concrete --B1PE50-- (see, e.g., Ref.~\cite{park2014computational}), which we take to be 50\% (C$_2$H$_2$)$_n$, 49\% concrete and 1\% boron carbide (B$_4$C).
We compute the flux of neutrons crossing a cylinder at a distance $\rho$ with $-1.5<z<1.5$~meters, which correspond to one of the dimensions of the detector volume.
This way, we can obtain more statistics by counting neutrons within an area $2\pi\rho \,\delta z$, where $\delta z = 3$~meters.
\begin{figure}[t]
\centering
\includegraphics[width=0.6\linewidth]{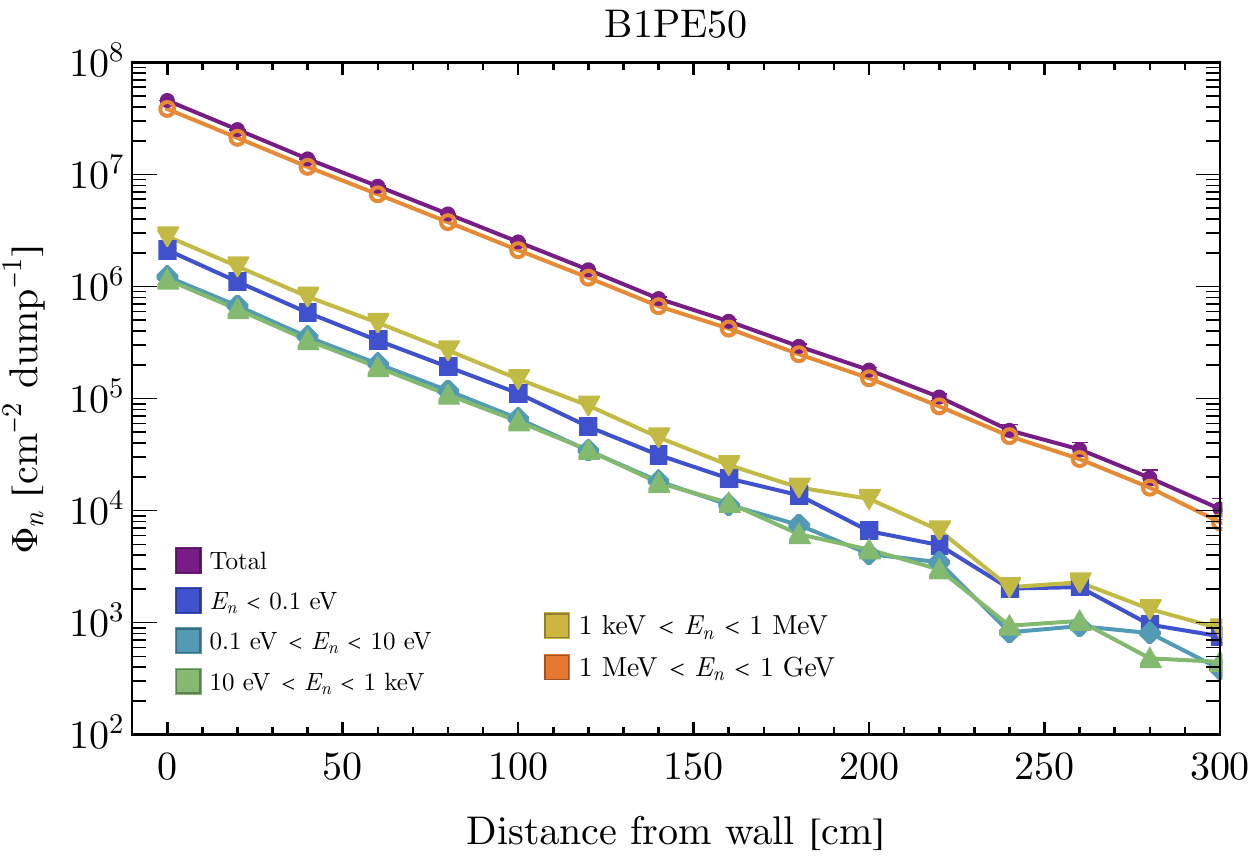}
\caption{\label{fig:neutron-background}Neutron flux for several ranges of neutron kinetic energy as a function of the distance from wall of the toy dump station setup (see text for details).\label{fig:NeutronFlux}}
\end{figure}

The neutron flux, as a function of the distance of the detector to the wall of the toy dump setup can be found in Fig.~\ref{fig:neutron-background}, for several neutron energy ranges.
We can see that the neutron flux right after the walls of the dump station are gigantic, almost at $10^8$ neutrons per cm$^2$ per dump.
Nevertheless, for every 3 meters of shielding, the flux drops by about 3.6 orders of magnitude.
Extrapolating to longer distances, a detector at 15 meters from the dump would have 9.5~meters of shielding and thus an expected neutron background around $2\times 10^{-4}/{\rm cm}^2/{\rm dump}$.
This neutron flux is much more manageable than the one at the wall.
Dedicated studies may be able to further suppress this background by considering different shielding material (e.g., iron to reduce the $E_n > 1~{\rm MeV}$ component of the flux).
Alternatively, one may push the detector location further from the dump by e.g. 3~meters, reducing the background to $5\times10^{-8}/{\rm cm}^2/{\rm dump}$ and losing only about 30\% of the neutrino flux.
Furthermore, the slow neutron component of the background could be mitigated by employing timing cuts: a neutron with $1$~eV of kinetic energy takes about $0.1$~ms to travel 1.4 meters. 

Another source of backgrounds not associated with the LHC Beam Dump operation comes from cosmic ray (CR) muons. 
This is particularly important for the detector technology we will consider later, liquid argon time projection chambers, due to the slow drift time.
We can perform a quick comparison to existing experiments to show that this background is not a problem in LEvEL.
Let us take as an example the 1~kton MINOS near detector, which is located in a cavern roughly 100~meters underground at the Fermilab site.
The rate of CR muons at MINOS near detector is 270~Hz~\cite{Habig:2005hc}~\footnote{Any muon not associated to a neutrino scattering event that enters the detector is considered a cosmic ray muon.}.
If we consider a similar rate at LEvEL (e.g. for a detector of roughly the size of MINOS near detector), we would have 0.023 CR muons per dump.
Even if we assume a LArTPC detector like MicroBooNE (89 ton fiducial mass), whose readout is of order 2~ms due to the drift time of electrons, this would still translate into 0.5 CR muons per readout.
MicroBooNE has shown that by considering the topological information on CR muon events and light collection, an overall CR muon rejection power of $7\times10^{-6}$ can be achieved~\cite{Abratenko:2021bzb}.
As we will see later, the rate of events for the processes of interest in a hypothetical LArTPC detector are much higher than $10^{-6}$ per dump.
Even if one considers a larger LArTPC, the CR muon rejection is still sufficient to make these backgrounds negligible --the ProtoDUNE detector, with a fiducial mass of 420 tons, is segmented in order to keep the drift time at the 2.2 millisecond level~\cite{Abi:2020mwi}.
In what follows, we will assume that beam related backgrounds can be sufficiently mitigated and that cosmic induced backgrounds are negligible due to the small duty factor. 

\section{Physics with Low-Energy Flux Perpendicular to Beam}\label{sec:OffAxis}

In this section, we discuss the physics processes that we can better understand by operating LEvEL at Position C, perpendicular to the incident proton beam direction. We study two specific neutrino-scattering processes in detail: low-energy $\nu_e$ scattering on argon (Section~\ref{subsec:SNApplications}) which is relevant for supernova neutrino studies, and \cevns (Section~\ref{subsec:CoherentApplications}) with a wide range of applications. We will demonstrate that measurements of these processes at LEvEL can provide important cross-section information to the broader neutrino community. Finally, in Section~\ref{subsec:BSM}, we briefly offer some remarks regarding searches for beyond-the-standard-model physics at LEvEL and discuss the weaknesses of such searches relative to those at other facilities.

%%%%%%%%%%%%%%%%%%%%%%%%%%%%%%%%%%%%%%%%%
\begin{table}[t]
\caption{Number of events per dump for the different interactions in a 100 ton liquid argon detector, except for CE$\nu$NS which we assume 1 ton. CC-LE corresponds to CC interactions with $E_\nu\leq 10$ GeV, CC-HE in the range $10~\GeV\leq E_\nu < 100$ GeV, and CC-VHE for $E_\nu \geq 100$ GeV. \label{tab:couplings}}
    \centering
    \begin{tabular}{ccccc}
        \toprule\toprule
        \multicolumn{5}{c}{Events per dump for 100 ton liquid argon detector (1 ton for CE$\nu$NS)}  \\ \midrule\midrule
        \multicolumn{2}{c}{Interactions} & Position A & Position B & Position C \\ \midrule\midrule
        \multicolumn{2}{c}{CE$\nu$NS (1~ton) } & 0.68 & 0.33 & 0.16\\ \midrule
        \multicolumn{2}{c}{$\nu_e$-Ar} & 2.0 & 1.8 & 1.6\\  \midrule
        \multicolumn{2}{c}{$\bar{\nu}_e$-Ar} & 0.019 & 0.018 & 0.010\\  \midrule
        \multicolumn{2}{c}{$\nu_x$-Ar} & 0.13 & 0.096 & 0.042\\  \midrule
        \multirow{3}*{CC-LE} & $\nu_e+\bar{\nu}_e$ & 3.1 & 0.24& 0.008\\  
        &$\nu_\mu+\bar{\nu}_\mu$ & 51 & 4.8 & 0.23\\  
        &$\nu_\tau+\bar{\nu}_\tau$& 0.016 & -- & -- \\  \midrule
       \multirow{3}*{CC-HE} & $\nu_e+\bar{\nu}_e$ & 2.0 & -- & --\\  
        &$\nu_\mu+\bar{\nu}_\mu$ & 13.0 & -- & --\\  
        &$\nu_\tau+\bar{\nu}_\tau$ & 0.24 & -- & --\\ \midrule
        \multirow{3}*{CC-VHE} & $\nu_e+\bar{\nu}_e$ & 1.9 & -- & --\\  
        &$\nu_\mu+\bar{\nu}_\mu$ & 3.0 & -- & --\\  
        &$\nu_\tau+\bar{\nu}_\tau$ & 0.3 & -- & --\\ \midrule
        \bottomrule
    \end{tabular}
\end{table}
%%%%%%%%%%%%%%%%%%%%%%%%%%%%%%%%%%%%%%%%%

To frame our discussion, in Table~\ref{tab:couplings}, we display the expected number of events of different types at Positions A, B, and C \textit{per beam dump}, assuming a 100 ton liquid argon detector for all processes except CE$\nu$NS, for which we take a 1 ton detector mass. 
These rates are calculated before any signal identification efficiencies are taken into account, however, we note that the event rates are fairly high. Specifically, at Position C, we can expect 1.6 $\nu_e$-Ar scattering events per beam dump for a detector with mass comparable to SBND~\cite{Machado:2019oxb}. 

\paragraph{Alternate Detector Considerations}
Throughout this work, we have centered our discussion on liquid argon detectors, motivated by currently unmeasured processes which are highly relevant for supernova detection, as well as the fact that many large-scale liquid argon detectors are currently in their planning stages. Regardless, the large, bunched neutrino flux near the LHC Beam Dump offers the possibility of performing neutrino-related measurements in other detector environments as well. 
For example, the $\bar\nu_e$ flux could be measured well with an inverse beta decay (IBD) detector, which could probe beyond standard model physics scenarios such as sterile neutrinos (similar to LSND~\cite{Aguilar:2001ty}) and leptonic non-unitarity. 
A water-based liquid scintillator, like the technology of the THEIA proposal, where a combined measurement of Cherenkov and scintillation lights enables for a better energy and direction reconstruction, could be used to measure IBD together with CC and NC interactions on oxygen that are relevant for Supernova detection~\cite{Askins:2019oqj}.
Investigating the physics potential of other detectors in this experimental setup is beyond the scope of this paper and may be pursued in a future work.

\subsection{Calibration of Supernova Detection Process}
\label{subsec:SNApplications}
Many next-generation neutrino experiments hope to be able to observe neutrinos from a galactic supernova during their lifetime. 
Several future experiments, notably the Jiangmen Underground Neutrino Observatory, aim to study interactions of $\overline{\nu}_e$ undergoing inverse beta decay, a well-understood process. 
In contrast, DUNE, the liquid argon Deep Underground Neutrino Experiment, offers a new, exciting way of studying supernova neutrinos via the process $\nu_e\,^{40}$Ar $\to e^-\,^{40}$K$^*$, with the excited potassium producing one or more photons when it decays to its ground state. 
This process has not yet been observed in a liquid argon time projection chamber\footnote{Recently, it has been suggested the possibility of having an additional detector at the SNS complex, as part of the plan to implement a Second Target Station, that could also measure the $\nu_e\,^{40}$Ar process~\cite{Scholberg:LOI}.}.
In order to extract information on neutrino physics out of a supernova detection, then, we must attain a better understanding of this cross section. Additionally, measurements of solar neutrinos in DUNE will rely on this process at lower neutrino energies ($E_\nu \approx 12$ MeV). Cross section uncertainties at the level of $\mathcal{O}(1\%)$ are important for DUNE to be able to use these solar neutrinos to improve our understanding of neutrino mixing~\cite{Capozzi:2018dat}.
%%%%%%%%%%    FIG     %%%%%%%
\begin{figure}[tb]
\centering
\begin{subfigure}[b]{.32\linewidth}
\centering
\includegraphics[width=\textwidth]{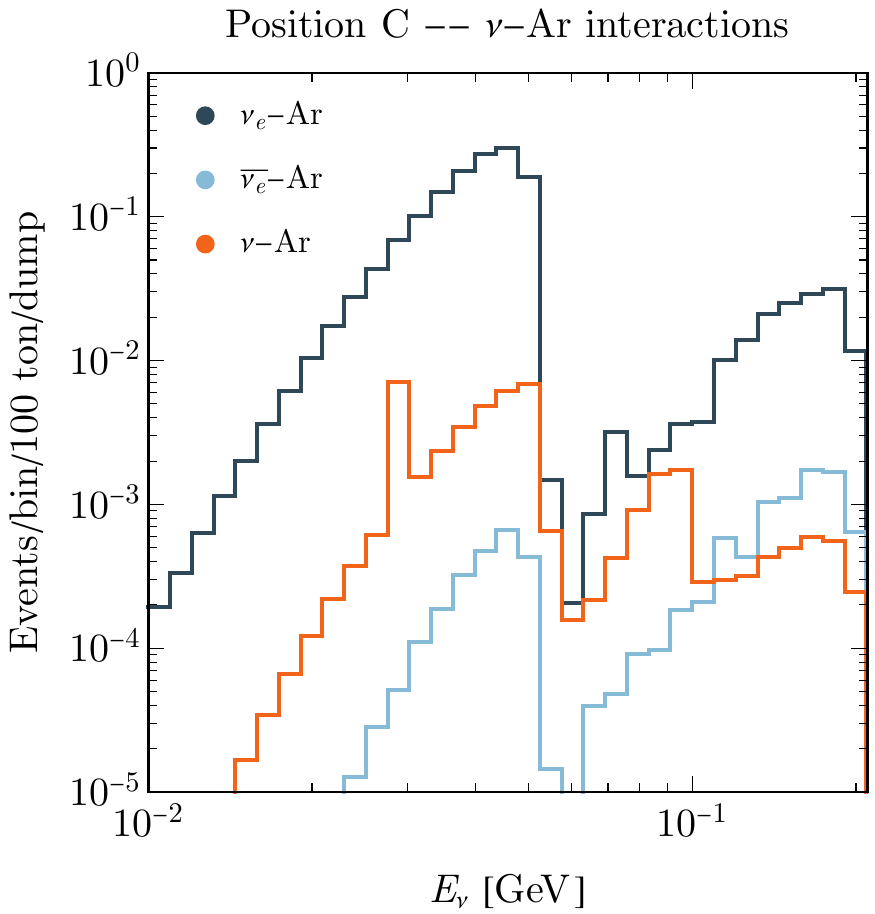}
\caption{Neutrino-Argon events. }\label{fig:1b}
\end{subfigure}
\begin{subfigure}[b]{.32\linewidth}
\centering
\includegraphics[width=\textwidth]{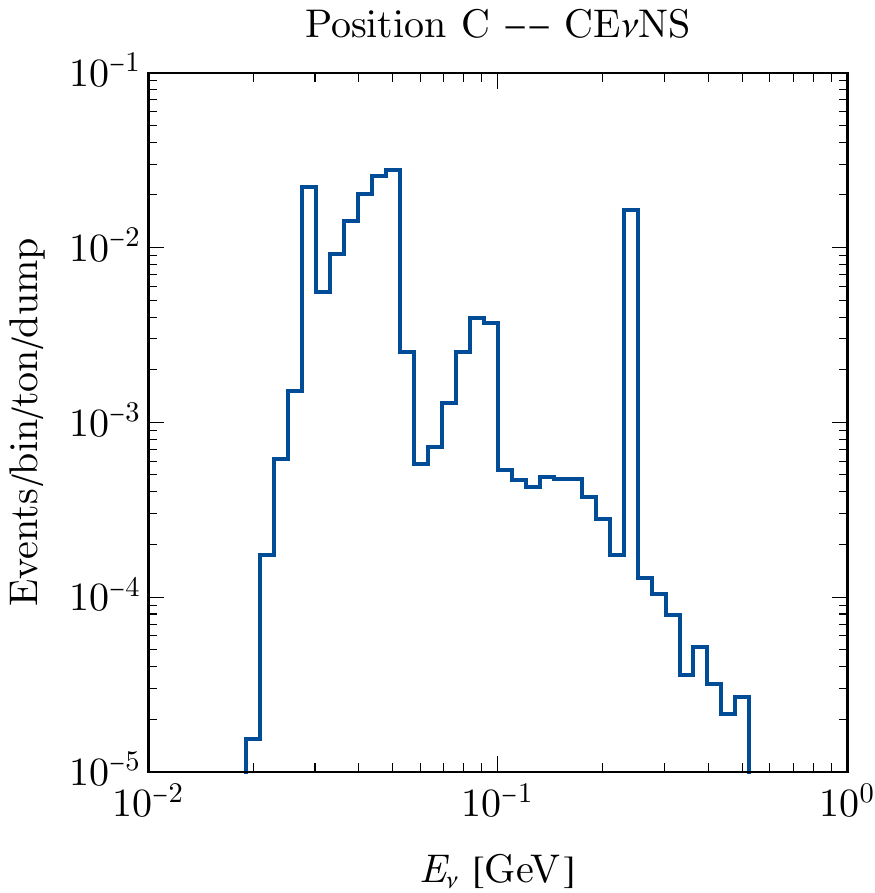}
\caption{Coherent Scattering events}\label{fig:1a}
\end{subfigure}\vspace{10pt}
\begin{subfigure}[b]{.32\linewidth}
\centering
\includegraphics[width=\textwidth]{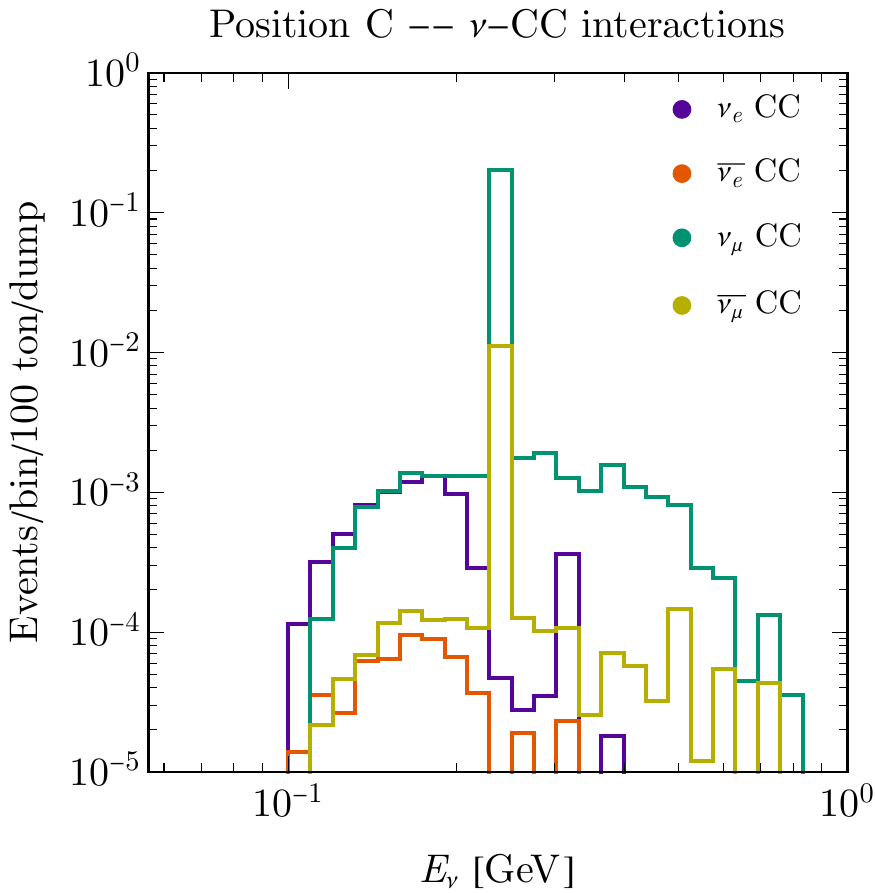}
\caption{Charged-current neutrino events}\label{fig:1c}
\end{subfigure}
\caption{Event spectra for different interaction types from the low-energy neutrino flux at Position C. The different event types we display are CE$\nu$NS (\subref{fig:1a}) ; neutrino-Argon interactions (\subref{fig:1b}) where the lines correspond to $\nu_e$ Ar (dark blue), $\overline{\nu}_e$ Ar (light blue), $\nu$ Ar Neutral Current (orange); and the four possible charged-current quasi-elastic processes (\subref{fig:1c}) with the lines corresponding to $\nu_e$ (purple), $\overline{\nu}_e$ (orange), $\nu_\mu$ (green), and $\overline{\nu}_\mu$ (dark yellow). All event rates are calculated per LHC beam dump and assume a detector volume of 100 ton (1 ton) for $\nu$-Ar and $\nu$-CC (CE$\nu$NS) scatterings.
\label{fig:EvtSpectrumPosC}}
\end{figure}
%%%%%%%%%%%%%%%%%%%%%%%%%%%%%%%%%%%%%%%%%

To study this process with actual data, as well as the related processes $\bar\nu_e\,^{40}{\rm Ar}\to e^+\, ^{40}{\rm Cl}^*$ and $\nu\,^{40}{\rm Ar}\to\nu\,^{40}{\rm Ar}^*$, we propose to deploy a $100$~ton liquid argon time projection chamber, similar to the ProtoDUNE detector, at Position C. We divide the remaining discussion into three pieces. First, the expected signal and its characteristics, as well as how we parameterize the cross section that we plan on measuring. Second, we discuss potential background contributions to an analysis of this type. Lastly, we discuss systematic uncertainties associated with this analysis and provide our projected measurement capabilities at LEvEL.

\subsubsection{Signal Characteristics and Cross Section}
\label{subsubsec:SNSignal}
 
Using the fluxes discussed cf Fig.~\ref{fig:FluxABC}, we determine the number of expected neutrino scattering events of different types and display them in the left panel of Fig.~\ref{fig:EvtSpectrumPosC}.
These event spectra give the expected number of events in each true neutrino energy bin assuming a $100$~ton detector at Position C, for each individual beam dump. 
We will focus first on the charged-current $\nu_e$-Ar events (dark blue), as it is the dominant contribution and this cross section has not been measured.

Early calculations of this cross section can be found in Refs.~\cite{Ormand:1994js, Bhattacharya:1998hc}.
We use the total cross section from Ref.~\cite{snowglobes}, which is similar to the one obtained using the MARLEY package~\cite{Gardiner:2018zfg,Gardiner:2021qfr}. This process results in one outgoing electron --- with kinetic energy approximately 30 MeV~\cite{Gardiner:2020ulp} --- and an excited potassium-40 atom. The de-excitation of $^{40}$K results in one or more photons being emitted, or, less often, a neutron being emitted. Ref.~\cite{Gardiner:2020ulp} discusses the spectrum of these de-excitation photons in some detail, including focusing on the spectrum of emission when the incoming $\nu_e$ are coming predominantly from muon decay-at-rest, like the situation considered here.

Our goal is to determine how well this cross section can be measured at LEvEL. 
To estimate this, we define a fit cross section which is essentially the exact cross section $\sigma(E_\nu)$ extracted from Ref.~\cite{snowglobes}, with a free normalization $R$ and a spectral index parameter $\gamma$, namely
\begin{equation}\label{eq:NuEXSec}
\sigma_{\rm fit}(\nu_e\ ^{40}\mathrm{Ar} \to e^-\ ^{40}\mathrm{K}^*) = R  \left(\frac{E_\nu}{E_0}\right)^\gamma \sigma(E_\nu).
\end{equation}
We fix the arbitrary parameter $E_0=40$~MeV to reduce the measurement correlation between $R$ and $\gamma$. Our goal then will be to determine how well a given exposure of data collection at LEvEL can measure $R$ and $\gamma$.

\subsubsection{Background Contributions to Charged-Current Scattering}
\label{subsubsec:SNBackgrounds}
As discussed in Section~\ref{subsubsec:SNSignal}, the signature of $\nu_e$ scattering on Argon results in an outgoing, tens-of-MeV electron and one or more MeV-scale photons. In Section~\ref{sec:Forward}, we discussed the fluxes of different particles that may contribute as background in different neutrino-scattering experiments near the LHC Beam Dump, namely muons and photons. At position C, where we plan on performing the $\nu_e$-Ar measurement, the muon flux is low enough to be discarded. 

On the other hand, the neutron flux is significant. In the discussion surrounding Fig.~\ref{fig:NeutronFlux}, we demonstrated that, at a distance of 15 m from the Beam Dump and with sufficient shielding, we expect a flux below $2\times 10^{-4}$/cm$^2$/dump at Position C. With a 100 ton detector, this implies $\mathcal{O}$(10) neutrons passing through the detector during each data collection period, compared with an expected $1.6$ charged-current scattering events expected during each window. At worst, every neutron will interact and deposit some energy during this data collection window. However, as discussed in great detail in Ref.~\cite{Capozzi:2018dat}, neutrons with ${\sim}$MeV energies (which make up the bulk of the flux at Position C) will predominantly be captured on liquid Argon, and this capture will cause photons to be released. These photons can mimic electrons with low kinetic energy, where the better the energy resolution of the detector, the lower energy they appear. For reasonable detector capability estimates, this signature will appear consistent with electrons with recoil energy well below 10 MeV. In contrast, our signal consists of electrons with 30 MeV recoil energies -- therefore these neutrons will provide negligible background in a search for $\nu_e + $ Ar $\to e^-$ $^{40}$K$^*$.

Finally, neutrino-induced neutrons (NINs) may be of concern. These are neutrons that arise from the incoming neutrino flux in/near the detector, where a neutrino scatters in someway that produces an outgoing neutron. Following the same arguments of the COHERENT collaboration~\cite{Akimov:2017ade,Akimov:2020pdx}, we expect that portion of the neutron flux will be very much subdominant compared against the prompt contribution considered above and in Fig.~\ref{fig:NeutronFlux}. For this reason, we disregard the NINs in this discussion.

\subsubsection{Systematic Uncertainties and Measurement Capability}
\label{subsubsec:SNSystMeasurement}
Using the expected event distribution shown in the left panel of Fig.~\ref{fig:EvtSpectrumPosC} along with the parameterized cross section (depending on $R$ and $\gamma$) in Eq.~\eqref{eq:NuEXSec}, we now set up our statistical analysis to measure this cross section. We assume that the energy resolution of the detector is powerful enough to divide events into the bins of Fig.~\ref{fig:EvtSpectrumPosC}(a), requiring ${\sim}2$ MeV capabilities. As argued in Section~\ref{subsubsec:SNBackgrounds}, this measurement is assumed to be background-free, given the disparity between the electron kinetic energies in our signal and the neutron-induced backgrounds.

We perform two separate types of analyses --- one in which we assume the only uncertainties are statistical, and one in which we assume that there are systematic uncertainties. As a demonstrative example, we consider systematics arising from the two largest sources of uncertainties: the overall neutrino flux and the kaon-to-pion ratio. The former will provide an overall scaling of the event rate, whereas the latter will vary the relative size of the two peaks as a function of $E_\nu$ exhibited in Fig.~\ref{fig:EvtSpectrumPosC}(a). We assume that the overall flux uncertainty is constrained at the 5\% level and that the kaon-to-pion production ratio is known at the 10\% level. Another source of systematic uncertainties corresponds to the LHC luminosity measurement. Nevertheless, beam monitors, like those in ATLAS and CMS, have measured the luminosity at $\sim 2\%$ percent~\cite{Aaboud:2016hhf,Sirunyan:2021qkt}, while, in the future, a precision $< 1.5\%$ is expected~\cite{CMS-PAS-FTR-18-011}. Moreover, our results do not rely on the bunch-by-bunch luminosity, but on the overall number of protons dumped. 
Hence, we expect that luminosity uncertainty will be a subleading effect for LEvEL.

\begin{figure}[t]
\begin{center}
\includegraphics[width=\linewidth]{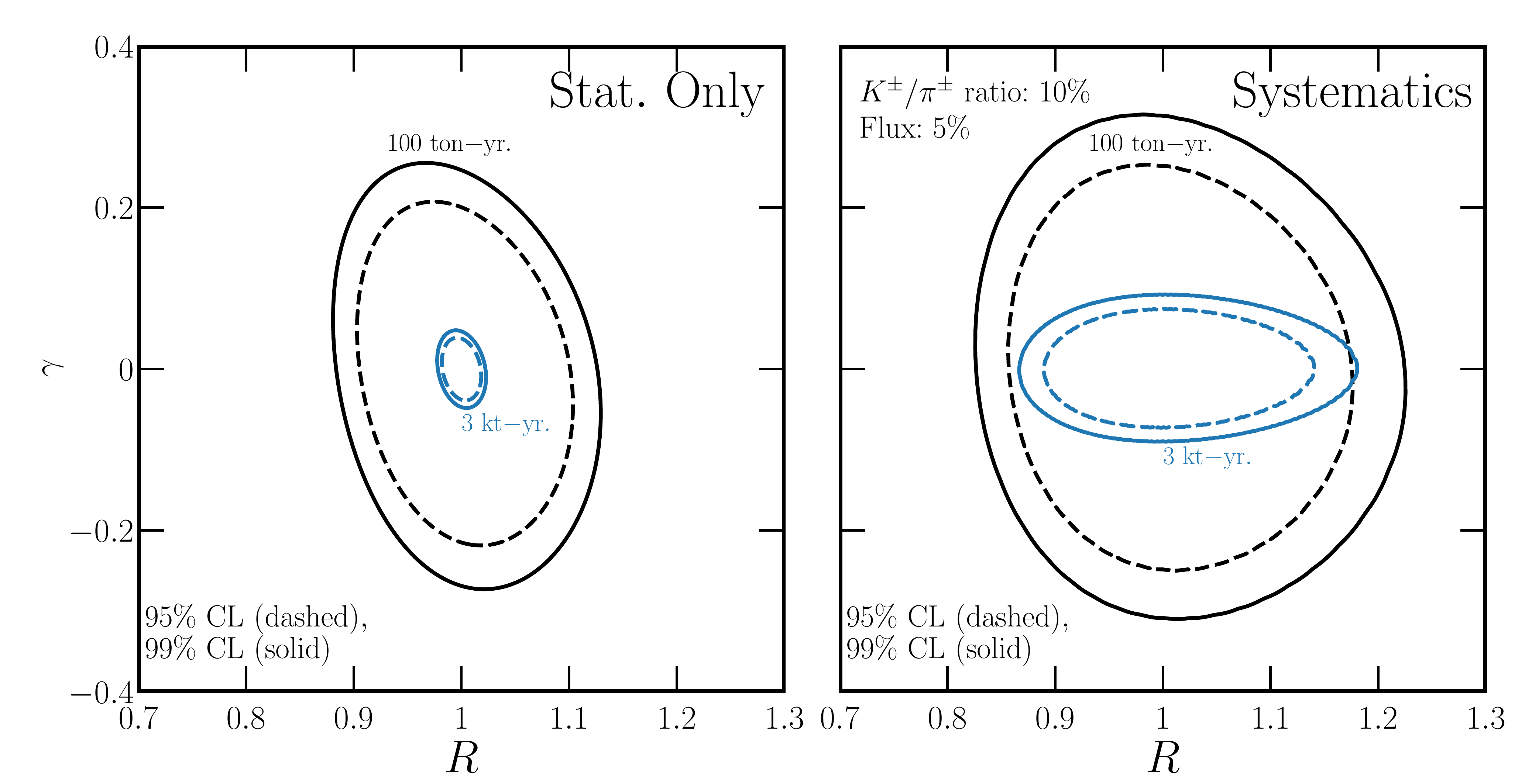}
\caption{Measurement capability of LEvEL for 100~ton-year (black) or 3~kton-year (blue) to measure the $\nu_e$ Ar cross section. We parameterize the cross section as given in Eq.~\eqref{eq:NuEXSec} with an overall normalization $R$ and a spectral tilt $\gamma$. Dashed (solid) curves correspond to 95\% (99\%) CL contours. The left panel assumes a statistics-only measurement, where the right panel assumes a 5\% correlated normalization uncertainty along with a $10\%$ uncertainty on the production ratio between charged kaons and pions.
\label{fig:NuEArXSec}}
\end{center}
\end{figure}

For both choices (without and with systematics), we perform the simulation assuming two different total exposures, $100$ ton-yr and $3$ kton-yr (for example, envisioning a larger, kiloton detector deployed for three years of data collection). Our results can be found in Fig.~\ref{fig:NuEArXSec} where the left (right) panel corresponds to the statistics-only (5\% systematics) analysis. The black (blue) contours correspond to the smaller (larger) exposure. In each case, dashed and solid lines correspond to $\Delta\chi^2=5.99$ and  $\Delta\chi^2=9.21$, respectively.

If systematic uncertainties can be completely avoided (left panel of Fig.~\ref{fig:NuEArXSec}), we can see that with an exposure of 3~kton-year, the normalization of this cross section could be measured to the $1\%$ level (at 68\% C.L.).
Even with only 100~ton-year of data collection, the normalization could be measured with $4\%$ precision.
Moreover, the spectral index $\gamma$ could be measured, at best, to the level of $\gamma \in [-0.02,0.02]$ with an exposure of 3~kton-year (or $[-0.09, 0.09]$ with 100~ton-year exposure). These measurements are naturally weaker when systematic uncertainties are included --- we summarize the measurement capability in these different scenarios in Table~\ref{tab:RGammaMeas}

\begin{table}[t]
\centering
\caption{Expected measurement capability of the normalization $R$ and spectral tilt $\gamma$ for the charged-current $\nu_e$ Ar scattering cross section, with different assumptions regarding exposure and systematic uncertainties. See text for more detail.\label{tab:RGammaMeas}}
\begin{tabular}{c||c|c||c|c}
& \multicolumn{2}{c||}{Statistics Only} & \multicolumn{2}{c}{With Systematics} \\ \hline
$1\sigma$ Range & $100$ ton-yr & 3 kt-yr & 100 ton-yr & 3 kt-yr \\ \hline\hline
Normalization ($R$) & $\pm4\%$ & $\pm 1\%$ & $\left[0.94, 1.07\right]$ & $\pm5\%$ \\ \hline
Spectral Tilt ($\gamma$) & $\pm 0.09$ & $\pm 0.02$ & $\pm 0.10$ & $\pm 0.03$ 
\end{tabular}
\end{table}

Of course, achieving these measurements in a realistic setting requires careful incorporation of the systematic uncertainties, with the right panel of Fig.~\ref{fig:NuEArXSec} demonstrating such impacts. Before concluding this portion of our analysis, we briefly offer some remarks on how these uncertainties may be understood and minimized. To reduce the neutrino flux uncertainty, we need to understand and/or measure the flux theoretically or monitor it via another experimental observable. In principle this could be done, for example, with 
$\nu_e$-electron scattering or a D$_2$O detector as recently proposed by the COHERENT collaboration~\cite{Rapp:2019vnv}.
For reference, a 15~ton-year D$_2$O detector exposure could constrain the neutrino flux at the 5\% level (statistical uncertainty only).
Another possibility would be to use the CCQE scattering of the monochromatic $\nu_\mu$ line from Kaon decay-at-rest~\cite{Nikolakopoulos:2020alk} as a handle on the kaon production, see Fig.~\ref{fig:EvtSpectrumPosC}(c). Such a measurement would also assist in constraining the relative production rates of kaons and pions. This systematic uncertainty is what hinders the measurement capability of $\gamma$ in Fig.~\ref{fig:NuEArXSec}(right).
The issue of monitoring and controlling the neutrino beam is common to all experiments whose goal is to measure unknown neutrino cross sections using pion and kaon decay-at-rest sources, and this includes the current proposal.

To conclude, we have demonstrated that, if systematic uncertainties can be controlled, or if the neutrino flux can be monitored, even 100~ton-year of data taking would allow us to obtain a deep  understanding of a critical neutrino-scattering process, $\nu_e~^{40}{\rm Ar}\to e^-\,^{40}{\rm K}^*$, while a large exposure of 3~kton-year would push this field to the precision era. 
Percent-level measurement of this cross section would allow us to extract detail information from any future supernova detection in a massive LArTPCs such as the DUNE far detector, as well as a deeper understanding of solar neutrinos in a similar enviornment.

\subsection{Applications to Coherent Scattering}
\label{subsec:CoherentApplications}

The low-energy coherent, elastic scattering process in which a neutrino of any flavor scatters off the target nucleus coherently, is a process we have only begun to understand, and has only been measured on a handful of target nuclei. One of these is argon~\cite{Akimov:2020pdx}. Future dark matter direct detection experiments will be limited in their dark matter searches by coherent scattering of incoming neutrinos from the atmosphere and the Sun~\cite{Billard:2013qya}. 
The impact of this so-called ``neutrino floor'' can be avoided or mitigated if we have a better understanding of the incoming neutrino flux and its coherent scattering cross section. 
Because several next-generation experiments will be composed of liquid argon, a precision understanding of this cross section as suggested here would serve to increase the sensitivity of these upcoming experiments.

In Section~\ref{subsec:SNApplications} we considered the possibility of having a 100~ton liquid argon time projection chamber at Position C.
Here, because we want to study the low-recoil coherent scattering process, we must imagine a smaller, more purpose-built detector. 
For that reason, in this subsection, we imagine a 1~ton liquid argon detector, built specifically for detecting CEvNS.
In terms of detector capabilities, we imagine that this is identical to the current COHERENT liquid argon detector, but with a larger mass. 
A 1~ton detector will allow for a relatively high event rate as seen in Fig.~\ref{fig:EvtSpectrumPosC}(b) and Table~\ref{tab:couplings}. As with Section~\ref{subsec:SNApplications}, we divide this analysis's discussion into parts devoted to the signal characteristics (and our cross section parameterization), background contributions, and the statistical/systematic analysis.

\subsubsection{Signal Characteristics and Cross Section}
\label{subsubsec:CohSignal}
Because the energy of the incoming neutrino cannot be precisely reconstructed in a coherent scattering event, we instead bin events according to the recoiling nuclear kinetic energy. This distribution is divided into $40$ keVnr bins (the kinetic energy of the recoiling nucleus is measured in units of keV of the nuclear recoil), consistent with the results of Ref.~\cite{Akimov:2020pdx}. While the spectrum rises for lower recoil energy, the reconstruction efficiencies drop significantly below $20$ keVnr, so we discard those events in our simulation. In Section~\ref{subsubsec:CohBackgrounds}, we will discuss how backgrounds contribute to this search and manifest in this observable.

The differential cross section for coherent neutrino scattering on a target nucleus can be expressed as~\cite{Freedman:1973yd}
\begin{equation}\label{eq:dSigmadT}
\frac{d\sigma}{dT_A} = \frac{G_F^2 m_A}{4\pi} \left( \mathcal{Q}_W^V\right)^2 \left[1 - \frac{T_A}{E_\nu} - \frac{m_A T_A}{2E_\nu^2}\right],
\end{equation}
where $T_A$ is the recoil kinetic energy of the final-state nucleus, $m_A$ is the nuclear mass, $E_\nu$ is the incoming neutrino energy, $G_F$ is Fermi's constant, and $\mathcal{Q}_W^V$ is the vector weak charge of the nucleus
\begin{equation}
\mathcal{Q}_W^V = \left[ g_V^p Z F_p(Q^2) + g_V^n N F_n(Q^2)\right].
\end{equation}
Here, $g_V^p = 4\sin^2\theta_W-1$ and $g_V^n = 1$ are the vector charges of protons and neutrons, with $\sin^2\theta_W=0.238$ being the Weinberg angle~\cite{Zyla:2020zbs}. $Z$ and $N$ are the number of protons and neutrons in the nucleus, and $F_x(Q^2)$ are the appropriate form factors of the nuclei, with $Q^2 = 2m_A T_A$ being the momentum transfer of the neutrino scattering process.
Using the Helm form factors for $F_x(Q^2)$~\cite{Engel:1991wq}, we can determine the overall neutrino scattering cross section as a function of $E_\nu$. 

Besides the cross section itself, there is another aspect that plays an important role in the detection of CEvNS, the \emph{quenching factor}.
When neutrinos  scatter off a nucleus coherently, only a fraction of the nuclear recoil energy turns into ionization energy, typically referred to as electron-equivalent energy.
This ratio between electron-equivalent and nuclear recoil energies is the quenching factor.
Experimentally, there is a minimum ionization energy for which a liquid argon detector will actually observe an event, and thus an uncertainty on the quenching factor will affect the detector's nuclear recoil energy threshold.
This in turn will translate into an uncertainty on the minimum neutrino energy that can lead to observable CEvNS events in the detector.
For example, the CENNS-10 detector in COHERENT~\cite{Akimov:2020pdx} experiences a sharp drop in its efficiency for electron-equivalent energies around 4-5~keVee (The quenching factor provides a conversion between keVnr and keVee, the `electron equivalent' energy).
The experimental determination of the quenching factor yields a quenching of about 25\% at these energies.
Therefore, the minimum nuclear recoil threshold for CENNS-10 is about 16-20~keVnr.
Such a minimum nuclear recoil energy can only be achieved by a minimum neutrino energy of $E_{\rm thr.}\simeq 18$~MeV.

For our purposes, it suffices to express the cross section and the quenching effect in terms of a few measurable parameters and analyze how well can we constrain these parameters.
We find that we can reasonably reproduce the energy dependence with four parameters and the equation
\begin{equation}\label{eq:cohXSec}
\sigma(\nu_x\ ^{40}\mathrm{Ar} \to \nu_x\ ^{40}\mathrm{Ar}) = R\times\sigma(E_0) \left(\frac{E_\nu}{E_0}\right)^{\gamma_1} \left(1 - \frac{E_{\rm thr.}}{E_\nu}\right)^{\gamma_2}\theta(E_\nu-E_{\rm thr.}),
\end{equation}
where, $R$ dictates the overall normalization with respect to the cross section at a reference energy $E_0$, which we take to be $100$ MeV. 
$\theta$ is a Heaviside step function.
Our simulations give, accounting for the Argon form factor, etc, $\sigma(E_0) \approx 5.3 \times 10^{-39}$ cm$^2$. 
The spectral index $\gamma_1$ dictates how this cross section grows with energy, and the combination of $E_{\rm thr.}$ and the spectral index $\gamma_2$ model the behavior of this cross section near threshold. 
This essentially accounts for the quenching uncertainties near the threshold.
For scattering on Argon, we find that we can reproduce the expected cross section fairly well for $\gamma_1 = 1.0$, $E_{\rm thresh.} = 18$ MeV, and $\gamma_2 = 3.0$.

As in Section~\ref{subsec:SNApplications}, our goal is to use the expected event rate to determine how well we can measure the corresponding cross section parameters with a certain amount of data collection.
However, due to the outgoing final-state neutrino, it is impossible to reconstruct the incoming neutrino energy 
on an event-by-event basis~\footnote{There has been efforts in developing CEvNS detectors with directionality, see e.g. Refs~\cite{Mayet:2016zxu, Battat:2016pap}. Detecting the direction of the recoiled nucleus along with its energy could, in principle, allow for a reconstruction of the incoming neutrino energy, thus enhancing the physics potential of our proposal.}. 
Instead, we take the expression in Eq.~\eqref{eq:dSigmadT} and determine the expected event spectra distributed as a function of the recoiling nuclear kinetic energy as our observable.

\subsubsection{Background Contributions to Coherent Scattering}
\label{subsubsec:CohBackgrounds}
The relatively large neutron flux at Position C is a source of backgrounds for coherent scattering. As discussed in Section~\ref{subsubsec:SNBackgrounds}, these ${\sim}$MeV neutrons can scatter off or be captured on argon and emit low-energy photons, which could be reconstructed as the signature of a recoiling argon nucleus in a CEvNS event. These types of backgrounds have been studied in great detail by the COHERENT collaboration~\cite{Akimov:2017ade,Akimov:2020pdx} in a situation where the relative neutron and neutrino fluxes are comparable to our proposal at Position C. Their signal and background rate contributions are roughly comparable, where the background exhibits a flatter distribution as a function of nuclear recoil energy than the small-recoil-peaked signal.

In our analysis, we assume that the total background rate is the same as the total signal rate, with the shape (as a function of recoil energy) of the background taken from Ref.~\cite{Akimov:2020pdx}. An alternate proposal is to move the detector from Position C an additional ${\sim}3$ meters further from the beam dump (at 18 m instead of 15 m, still perpendicular from the incident proton beam direction). With 3 m of additional B1PE50 shielding, the neutron flux can be lowered by a factor of more than $10^{3}$, resulting in a optimistic, nearly background-free search.\footnote{For idealistic results that consider the alternate proposal of a detector 18 m from the LHC Beam Dump, resulting in approximately zero beam-related neutron backgrounds, see Appendix~\ref{app:CEvNSZeroBackground}.} This additional distance would cause the isotropic neutrino flux to be about 70\% of the flux at 15~m, and the lower resulting statistics could be recovered by collecting data for a correspondingly longer period. Consistent with Ref.~\cite{Akimov:2020pdx}, we assume that the NIN contribution to the neutron flux is very much subdominant to the prompt one. We will study both of these contributions in greater detail in a future, dedicated study. We find that the measurement capability in this idealistic alternate scenario is comparable to what we obtain with a larger, $\mathcal{O}(1)$ background-to-signal ratio.

\subsubsection{Systematic Uncertainties and Measurement Capability}
\label{subsubsec:CohSystMeasurement}

We assume that the same systematics associated with the neutrino flux discussed in Section~\ref{subsubsec:SNSystMeasurement} will apply in a search for coherent scattering, specifically an overall normalization uncertainty of $5\%$ and an uncertainty on the kaon-to-pion production ratio of $10\%$. In contrast to the previous analysis, these systematic uncertainties do not have a significant impact on our coherent-scattering results as the cases we explore are all statistically limited. We have performed studies assuming both a 1 ton-yr and a 30 ton-yr (imagining a larger, 10 ton-scale detector operating for 3 years) exposures.

\begin{figure}[th]
\begin{center}
\includegraphics[width=0.7\linewidth]{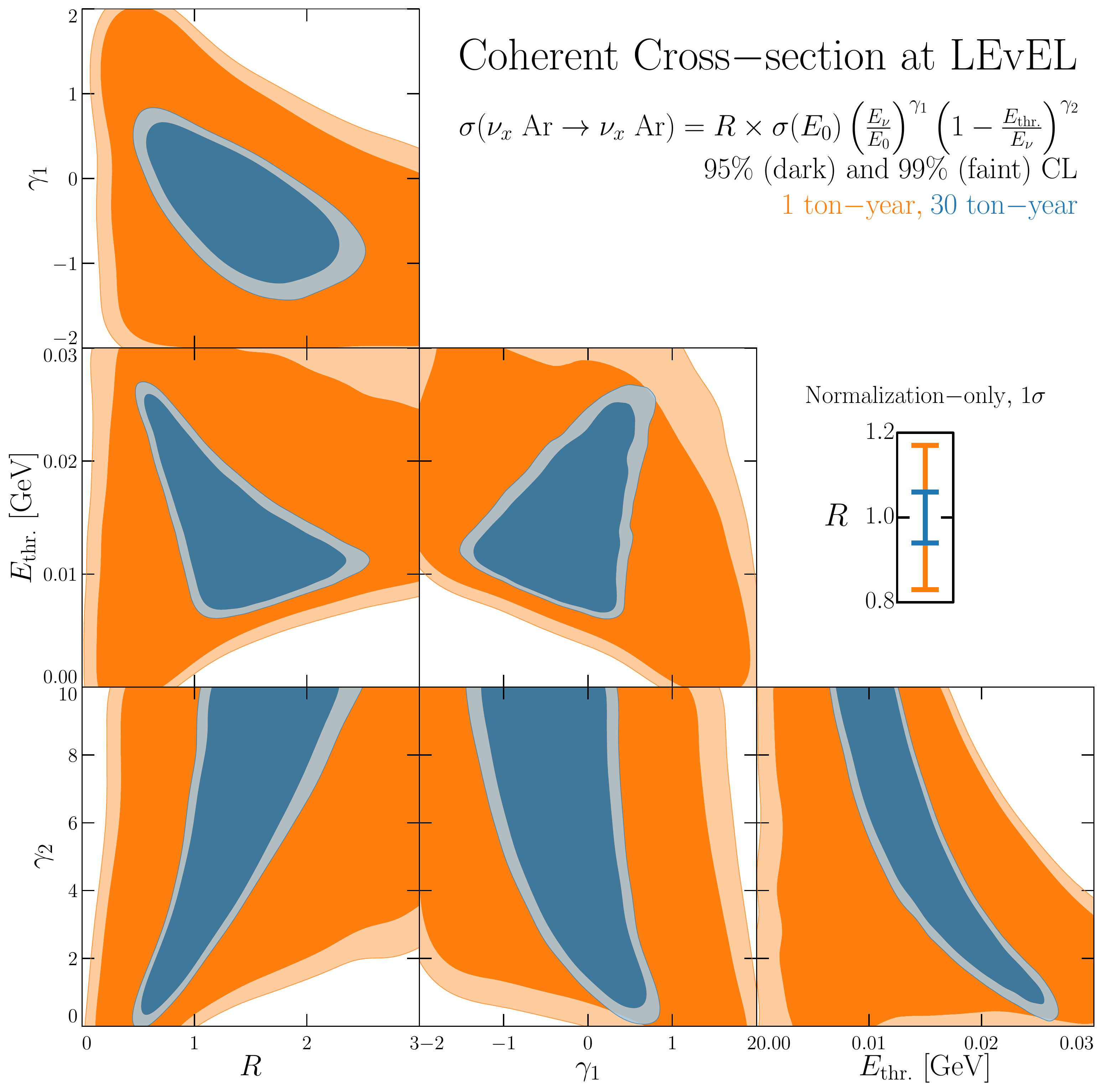}
\caption{Measurement capability of LEvEL for the coherent scattering cross section on argon assuming a signal-to-background ratio of 1 and systematic uncertainties (5\% overall flux uncertainty and 10\% kaon-to-pion production ratio uncertainty) marginalized over. Both a 1 ton-yr (orange) and a 30 ton-yr (blue, assuming a 10 ton-scale detector operating for 3 years) are considered. Here we parameterize this cross section as given in Eq.~\eqref{eq:cohXSec} and given in the top-right of this figure. For both exposures, we display 95\% (darker colors) and 99\% (fainter) credible regions, as extracted using the Bayesian inference tool {\texttt{PyMultiNest}}~\cite{Feroz:2013hea,Buchner:2014nha}. The inset displays the extracted $\pm1\sigma$ measurement of $R$ when only constraining the cross-section normalization, i.e. when $\gamma_{1}$, $\gamma_{2}$, and $E_{\rm thr.}$ are fixed.
\label{fig:NuCohXSec}}
\end{center}
\end{figure}

The resulting measurement capability of $(R,\ \gamma_1,\ E_{\rm thresh.},\ \gamma_2)$ is shown in Fig.~\ref{fig:NuCohXSec}. 
As one can see, there are large degeneracies in the determination of these parameters. 
The reason for this can be traced back to the impossibility of reconstructing the incoming neutrino energy from nuclear recoil energy.
The dependence of the nuclear recoil spectrum on the differential cross section is, at best, weak.
A more realistic goal for this experimental setup would be to determine the total flux weighted cross section.
We present that in the inset of Fig.~\ref{fig:NuCohXSec} labeled ``Normalization-only,'' which shows that a precision of 17\% or 6.0\% would be achievable for exposures of 1~ton-year and 30~ton-year, respectively. The equivalent of Fig.~\ref{fig:NuCohXSec} is presented in Appendix~\ref{app:CEvNSZeroBackground} as Fig.~\ref{fig:NuCohXSec_NoBkg} where we optimistically assume a slightly further detector distance (18 m) resulting in zero background. In this ideal, practically zero-background analysis, we find a normalization-only measurement\footnote{For comparison, we have also performed the equivalent normalization-only measurements when systematic uncertainties are not included. The resulting capability, when the signal-to-background ratio is 1, is $16\%$ ($2.9\%$) for 1 ton-yr (30 ton-yr) exposure. When the backgrounds are reduced to zero with 3 m additional shielding, the capability is $15\%$ ($2.7\%$) for 1 ton-yr (30 ton-yr) exposure.} capability of $16\%$ ($6.0\%$) for a 1 ton-yr (30 ton-yr) exposure, comparable to the results here with backgrounds.

Another way to visualize this measurement is to 
take the allowed region of $(R,\ \gamma_1,\ E_{\rm thresh.},\ \gamma_2)$ at a given confidence level and to determine the range of $\sigma(\nu_X\ ^{40}\mathrm{Ar}\to\nu_X\ ^{40}\mathrm{Ar})$ as a function of neutrino energy that this measurement allows, including the detector threshold and quenching effects.
\begin{figure}[!ht]
\begin{center}
\includegraphics[width=0.8\linewidth]{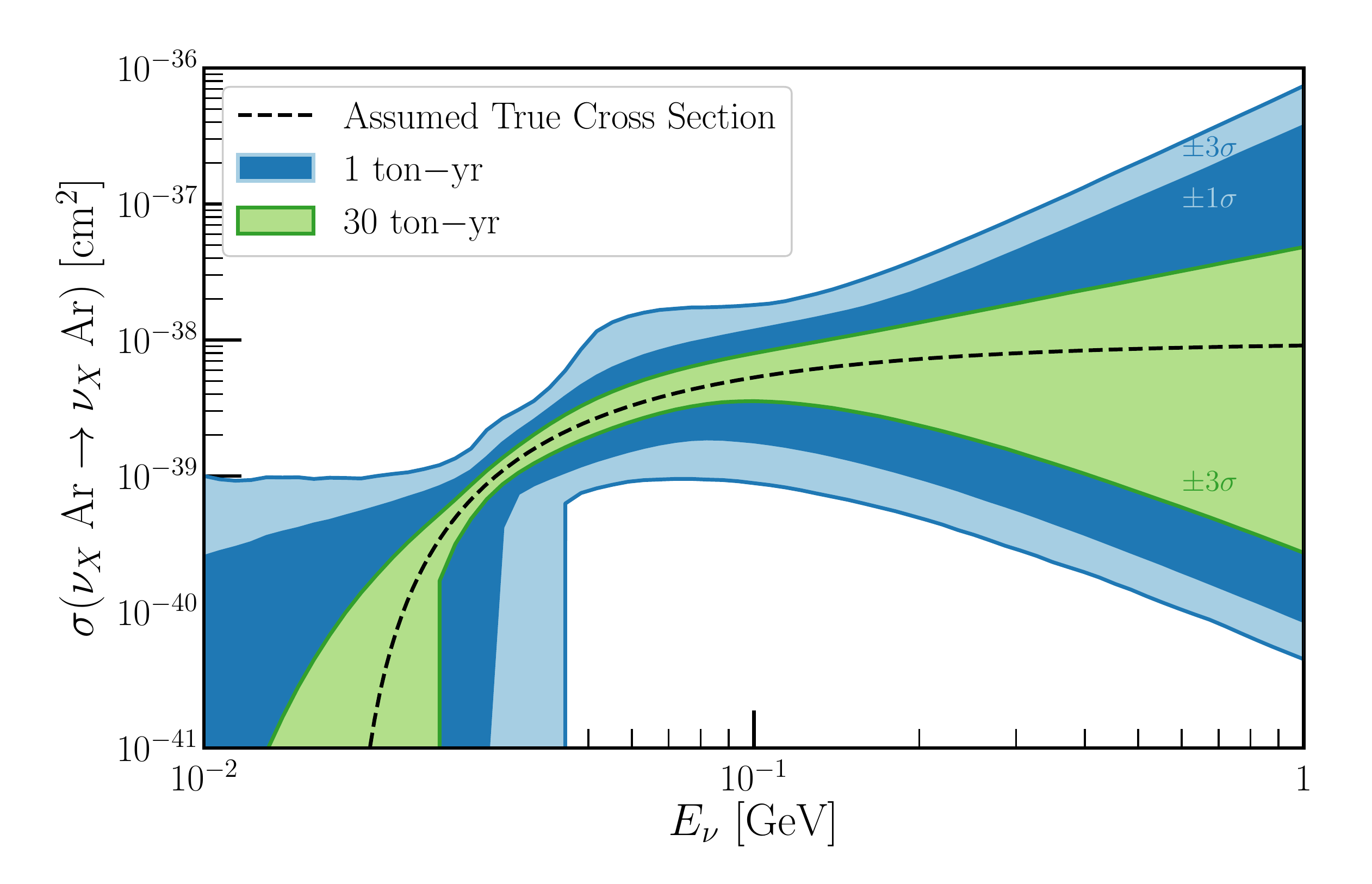}
\caption{Allowed range of neutrino coherent cross sections assuming 1 ton-yr (blue) and 30~ton-yr (green) exposures and using the coherent cross-section parameterization given in Eq.~\eqref{eq:cohXSec}. 
The cut-off at low neutrino energies models the detector threshold and the effect of quenching.
We determine the $\pm n\sigma$ range of this cross section by allowing the parameters $\left\lbrace R, \gamma_{1}, E_{\rm thr.}, \gamma_{2}\right\rbrace$ to vary within their joint $n\sigma$ allowed parameter space and finding the minimum/maximum cross section for all $E_\nu$. Here, systematic uncertainties ($5\%$ overall flux uncertainty and $10\%$ kaon-to-pion production ratio uncertainty) are marginalized over, and a signal-to-background ratio of $1$ is assumed.
\label{fig:CohXSecAllowed}}
\end{center}
\end{figure}
The resulting allowed cross section at $1$ and $3\sigma$ confidence is shown in Fig.~\ref{fig:CohXSecAllowed}, where we have taken the 1~ton-yr exposure as our benchmark. 
For comparison, we also show the $3\sigma$ allowed region for a larger exposure of 30~ton-yr (in which we envision a 10~ton-scale detector collecting data for 3 years).
Note that this measurement is not exactly constraining the allowed cross section over the entire energy range -- this is the allowed range as dictated by the parameterization we adopted in Eq.~\eqref{eq:cohXSec} given the measurements of the four free parameters. 
We see here that the main power in this measurement is coming from the incident neutrinos with $30$ MeV $\lesssim E_\nu \lesssim 60$ MeV, which is the energy range for the bulk of coherent-scattering events as seen in Fig.~\ref{fig:EvtSpectrumPosC}. Whereas Fig.~\ref{fig:CohXSecAllowed} includes both systematic uncertainties ($5\%$ overall normalization uncertainty and $10\%$ kaon-to-pion production ratio uncertainty) and a signal-to-background ratio of $1$, we also repeat this process optimistically assuming no background (and a slightly smaller per-year exposure) in Appendix~\ref{app:CEvNSZeroBackground} cf Fig.~\ref{fig:CohXSecAllowed_NoBkg}.

\subsection{Considerations on Searches for Beyond-the-Standard-Model Physics}\label{subsec:BSM}
Briefly, we wish to discuss the prospects of some types of beyond-the-Standard-Model (BSM) searches in the LEvEL setup. 
BSM searches that can be performed in proton- and electron-beam-dump environments vary from signatures of new particles scattering in a detector, or decaying within, among others. Searches for new-physics particles decaying (see, e.g., Refs.~\cite{Ariga:2018uku,Beacham:2019nyx,Berryman:2019dme}) in an active volume can often be (nearly) background-free, given the distinct signature relative to scattering-related backgrounds\footnote{As an example, let us consider the heavy neutral lepton scenario. We can have a new particle $N$ produced in the decays of charged kaons, then travel to the detector, and decay $N \to \mu^\pm \pi^\mp$. The signature then is muon/pion pairs, all with a common invariant mass $m_{\mu\pi}^2 = m_N^2$.}. 
This implies that a search proposal's strength is limited by the signal production rate, which is in turn limited by the production rate of the new-physics particle(s). In the cases we have explored, including searches for dark photons, dark Higgs bosons, axion-like particles, and heavy neutral leptons, we have found that the production rate at LEvEL is significantly lower than in other experimental setups, making such searches here less powerful.

If an experimental search for a BSM signature is \textit{not} zero-background, i.e. background rejection is important, LEvEL may be able to offer unique capabilities with the low duty factor we have discussed throughout. This includes, for instance, searches for light dark matter being produced associated with the beam and scattering off electrons or nuclei in a detector (see Ref.~\cite{Batell:2021blf} for a proposal of this search in the forward region at the LHC). We leave these studies for future work, but note that future searches at LEvEL may be complementary to those proposed at Fermilab for the next generation of neutrino-related experiments~\cite{Toups:LOI}.

\section{Discussion \& Conclusions}\label{sec:Conclusions}

The next generation of neutrino experiments aims to uncover much about the nature of neutrinos. 
This ranges from properties that impact neutrino oscillations to their many ways of interacting with matter. 
Further, if a nearby supernova occurs when these detectors are online, we can extract invaluable information of stellar evolution and neutrino properties by detecting supernova neutrinos.
All of these goals rely on a precise understanding of neutrino interactions.

In this work, we have demonstrated that a liquid argon neutrino detector, \textbf{LEvEL}, operating near the Large Hadron Collider beam dump, is capable of performing precise measurements of specific neutrino interaction processes to the \emph{level} required by next-generation experiments. 
While a detector in the forward region would offer the ability to measure ${\sim}$TeV neutrinos coming out of the beam dump, the muon and neutron backgrounds for such an experiment are very large.
An in-depth background mitigation study, employing magnetic fields to deflect the muons and passive shielding to reduce the neutron flux, would be needed to assess the feasibility of such proposal.
In contrast, a detector $20$~m from the beam dump but perpendicular to its direction can have very large neutrino signals from meson decay-at-rest and muon decay-in-flight processes with relatively small background. 
Moreover, since the LHC beam could be dumped twice a day for 86~$\mu$s each time, the large ratio between beam-off and beam-on operation allows for background reduction at a rate unseen by similar spallation sources of low-energy neutrinos.

We have considered two potential neutrino detectors at this location, each well-suited to measure an interaction process of interest to the future physics programme. 
A $100$~ton liquid argon time projection chamber (similar to SBND) is ideal for measuring the process $\nu_e~^{40}$Ar $\to e^-~^{40}$K$^*$, with the excited potassium decaying to its ground state and emitting a detectable photon. 
This process is of key importance to measuring the  $\nu_e$ flux coming out from a core collapse supernova explosion, especially in a large liquid argon time projection chamber like DUNE. 
The second detector we considered is a 1~ton liquid argon detector designed for detection of low-energy nuclear recoils, effectively a larger version of the CENNS-10 detector used by the COHERENT collaboration. 
This would allow for a percent level measurement of the coherent elastic neutrino-nucleus cross section.

Regarding BSM scenarios, we have determined that  promising searches would be those limited by backgrounds, instead of those that require high intensity.
LEvEL offers the highest background rejection factor among all experiments with beam power above a kilowatt. 
Future studies identifying these scenarios would be desirable.
Besides, new physics that modify neutrino cross sections, such as nonstandard interactions, could also profit from this proposal.

Finally, we would like to emphasize that the experimental setup proposed here is not unique to the LHC, and could be implemented in future high energy/high luminosity colliders, such as a 100~TeV proton-proton machine.
Building a future collider, having in mind an associated low energy/low background neutrino program could also enhance the capabilities of a setup like LEvEL.
Combining collider and neutrino efforts under a single umbrella will bolster the physics output of both programs.

\begin{comment}
%%%%%%%%%%    FIG     %%%%%%%
\begin{figure}[t]
\centering
\includegraphics[width=\textwidth]{Plots_a.pdf}
\caption{Events spectra for low (high) energy neutrinos in the left (right) top panels computed for a LAr detector at Position A. The bottom panel corresponds to the neutron, muon and photon fluxes for the same detector position. The right-bottom panel shows the total number of muons per ${\rm cm^2}$ as function of the distance from the interaction point.}
\label{fig:PA}
\end{figure}
%%%%%%%%%%%%%%%%%%%%%%%%%%%%%%%%%%%%%%%%%

%%%%%%%%%%    FIG     %%%%%%%
\begin{figure}[t]
\centering
\includegraphics[width=0.5\textwidth]{Plots_b.pdf}
\caption{Events spectra for low energy neutrinos in the top panels computed for a LAr detector at Position B. }
\label{fig:PB}
\end{figure}
%%%%%%%%%%%%%%%%%%%%%%%%%%%%%%%%%%%%%%%%%
\end{comment}

%%%%%%%%%%%%%%%%%%%%%%%%%%%%%%%%%%%%%%%%%%%%%%%%%%%%%%%%%%%%%%%%%%%%%%%%%%%%%%%%%%%%%%%%%%%%%%%%%%%%%%%%
\section*{Acknowledgements}
We are particularly grateful to Marco Calviani, Joachim Kopp, Anton Lechner and Bryce Littlejohn for invaluable comments on the draft.
We would also like to thank Francesco Cerutti, Shirley Li,  Jos{\'e} Maria Martin Ruiz and the users of the {\tt FLUKA} User Forum. 
Fermilab is operated by the Fermi Research Alliance, LLC under contract No. DE-AC02-07CH11359 with the United States Department of Energy. 
This project has received support from the European Union’s Horizon 2020 research and innovation programme under the Marie Skłodowska-Curie grant agreement No 860881-HIDDeN.
%%%%%%%%%%%%%%%%%%%%%%%%%%%%%%%%%%%%%%%%%%%%%%%%%%%%%%%%%%%%%%%%%%%%%%%%%%%%%%%%%%%%%%%%%%%%%%%%%%%%%%%%

\appendix

\section{Particle fluence and further background studies in the LHC beam dump}\label{ap:pflu}

In Sections~\ref{sec:Setup} and~\ref{sec:Forward}, we discussed the neutrino flux as well as the muon- and neutron-related backgrounds surrounding the LHCBD. 
In this Appendix, we provide further details on the neutrino flux and such backgrounds.
As we have seen in Fig.~\ref{fig:FluxABC}, the high-energy component of the neutrino flux is strongly reduced when going from Position A to B and completely gone in Position C. 
As this component could be very useful to perform studies of neutrino cross sections, such as deep inelastic scattering, and aspects of parton distribution functions, we further quantify the angular dependence of the high energy neutrino flux.

\begin{figure}[t]
\centering
    \includegraphics[width=0.46\textwidth]{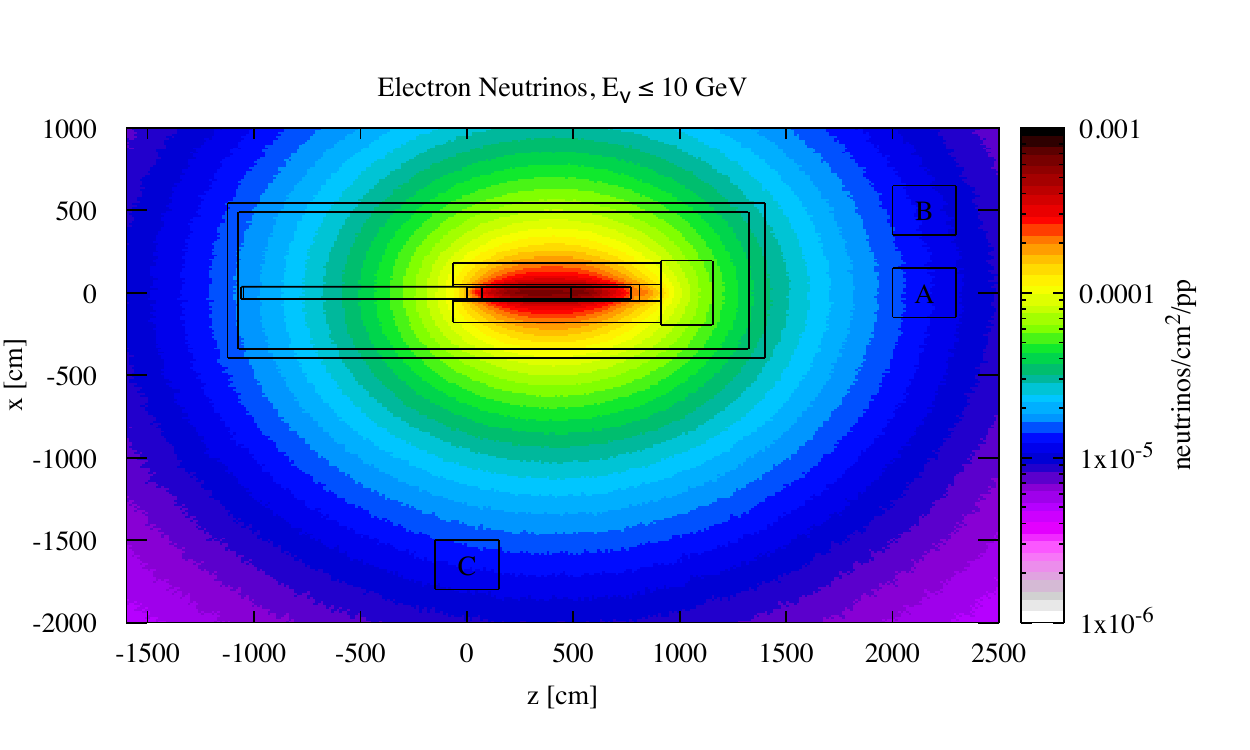}\hspace{0.5cm}
    \includegraphics[width=0.46\textwidth]{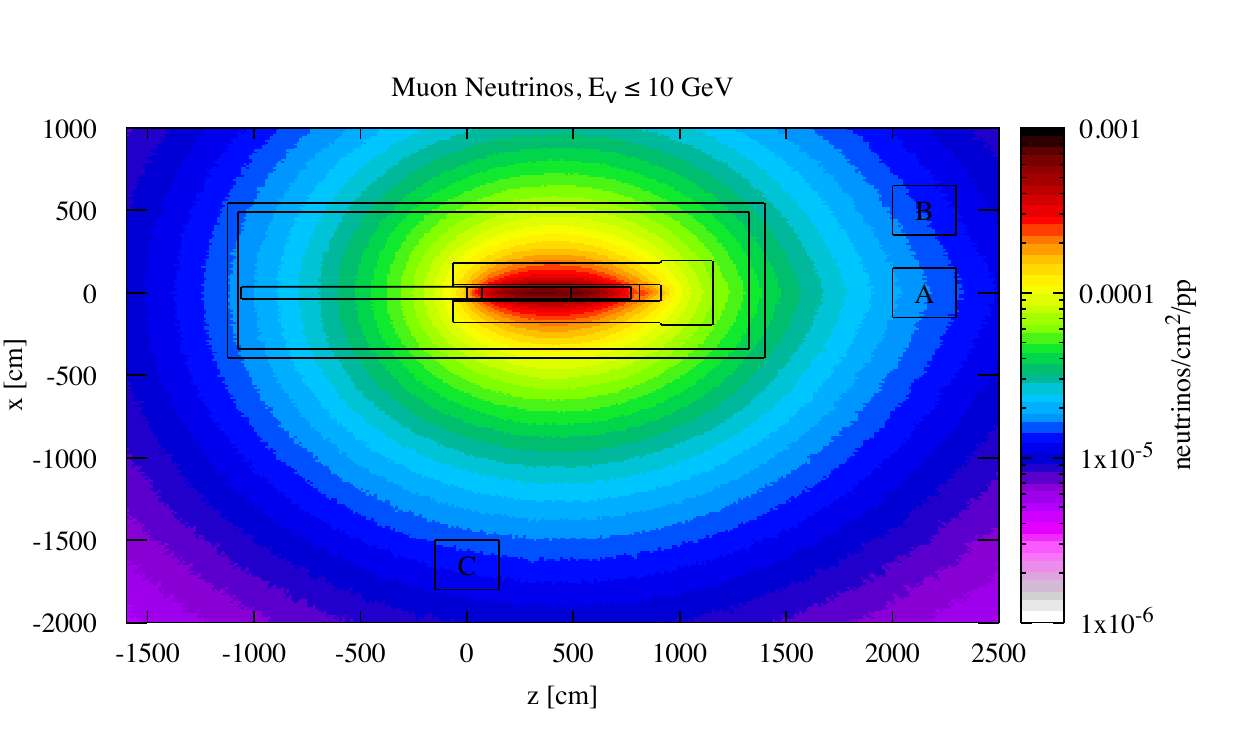}\\
    \includegraphics[width=0.46\textwidth]{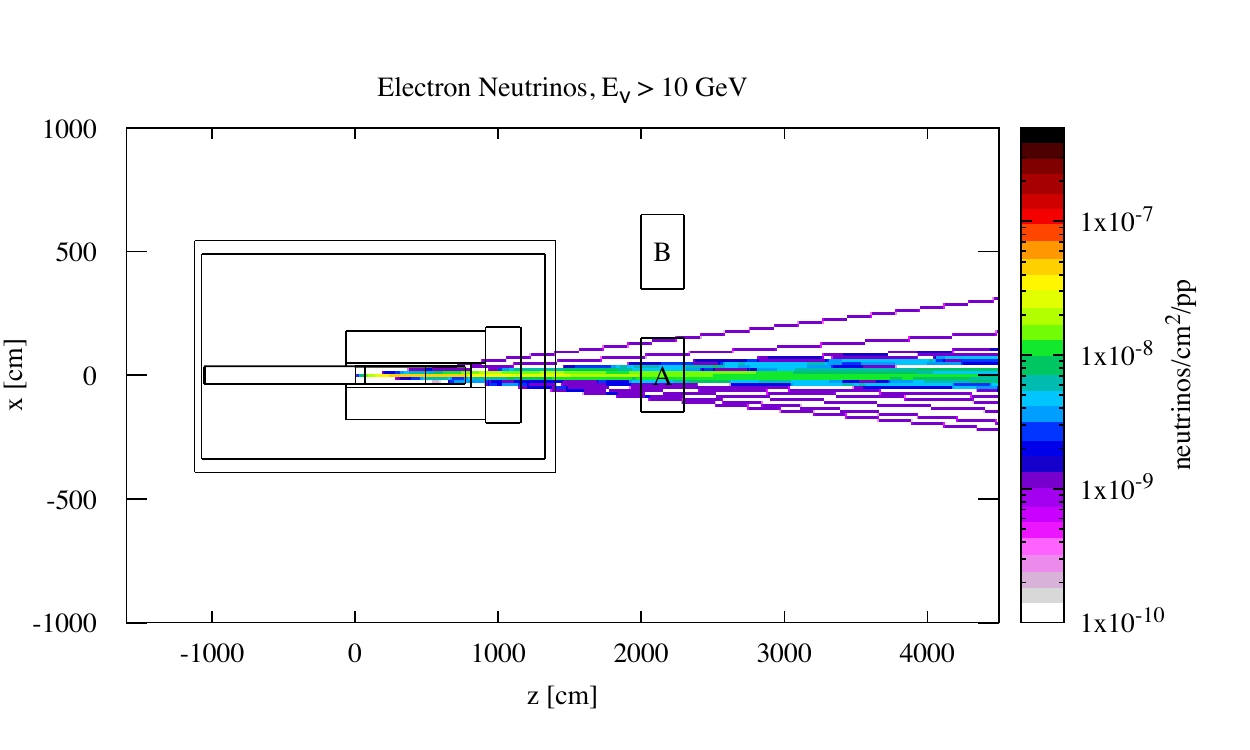}\hspace{0.5cm}
    \includegraphics[width=0.46\textwidth]{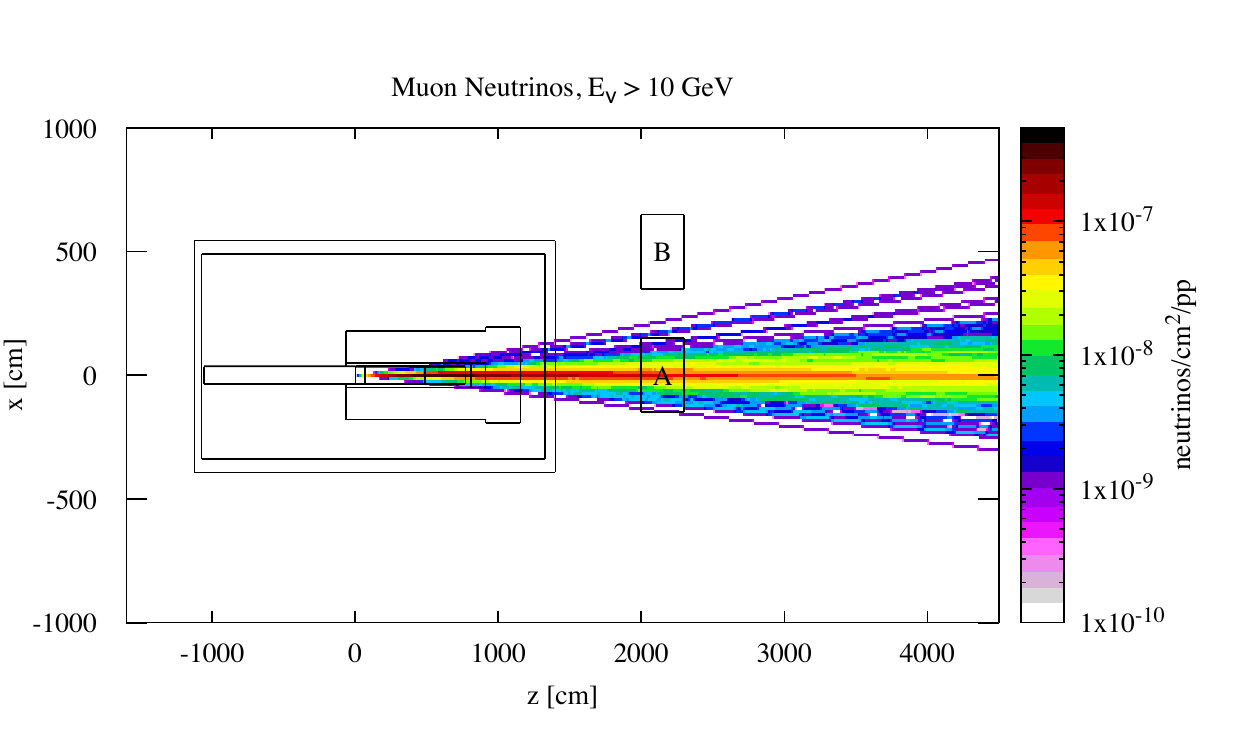}
    \caption{Neutrino fluence in ${\rm cm}^{-2}$ per primary proton for $E_\nu < 10~\GeV$ (top) and $E_\nu \geq 10~\GeV$ (bottom) for  $\nu_e$ (left) and $\nu_\mu$ (right). The isotropic component comes from mesons decaying at rest in the beam dump, while high energy neutrinos are mostly in the forward direction.}
    \label{fig:neutr}
\end{figure}

In Fig.~\ref{fig:neutr}, we show the two dimensional projection of the neutrino fluence per dump for each neutrino flavor.
Clearly, we can see that most neutrinos are produced in the dump and the fluence outside the dump follows a $1/L^2$ intensity drop, see upper panels. 
Nevertheless, inspecting the bottom panels, we find that the high energy neutrino component is very forward, as expected.
This can be see very clearly in the case of muon neutrinos (right) as these originate in 2-body decays of high energy pions.
Electron (anti)neutrinos are typically produced in 3-body decays, exhibiting a less forward flux.

%%%%%%%%%%    FIG     %%%%%%%
\begin{figure}[t]
\centering
\includegraphics[width=0.45\textwidth]{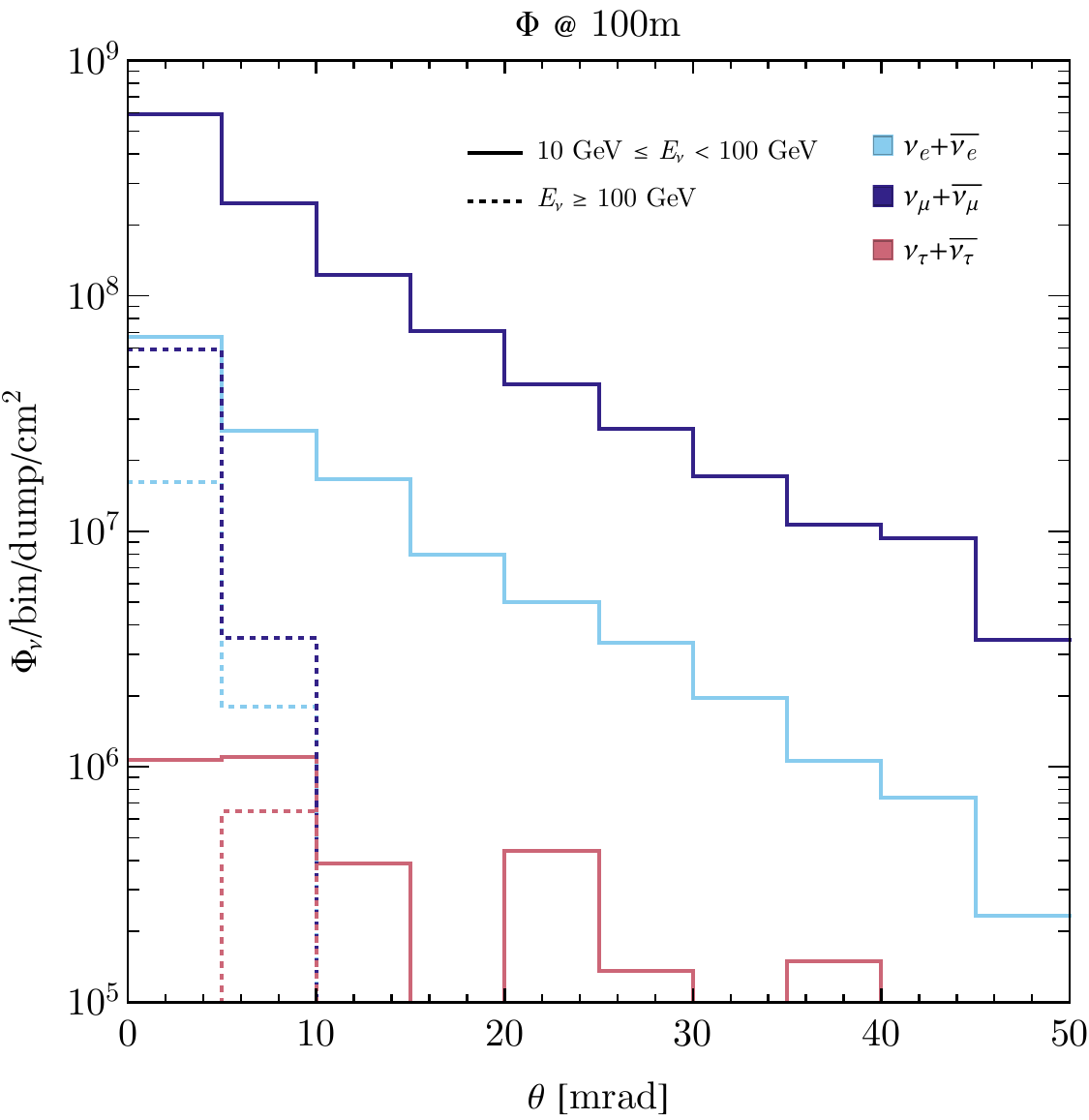}
\includegraphics[width=0.45\textwidth]{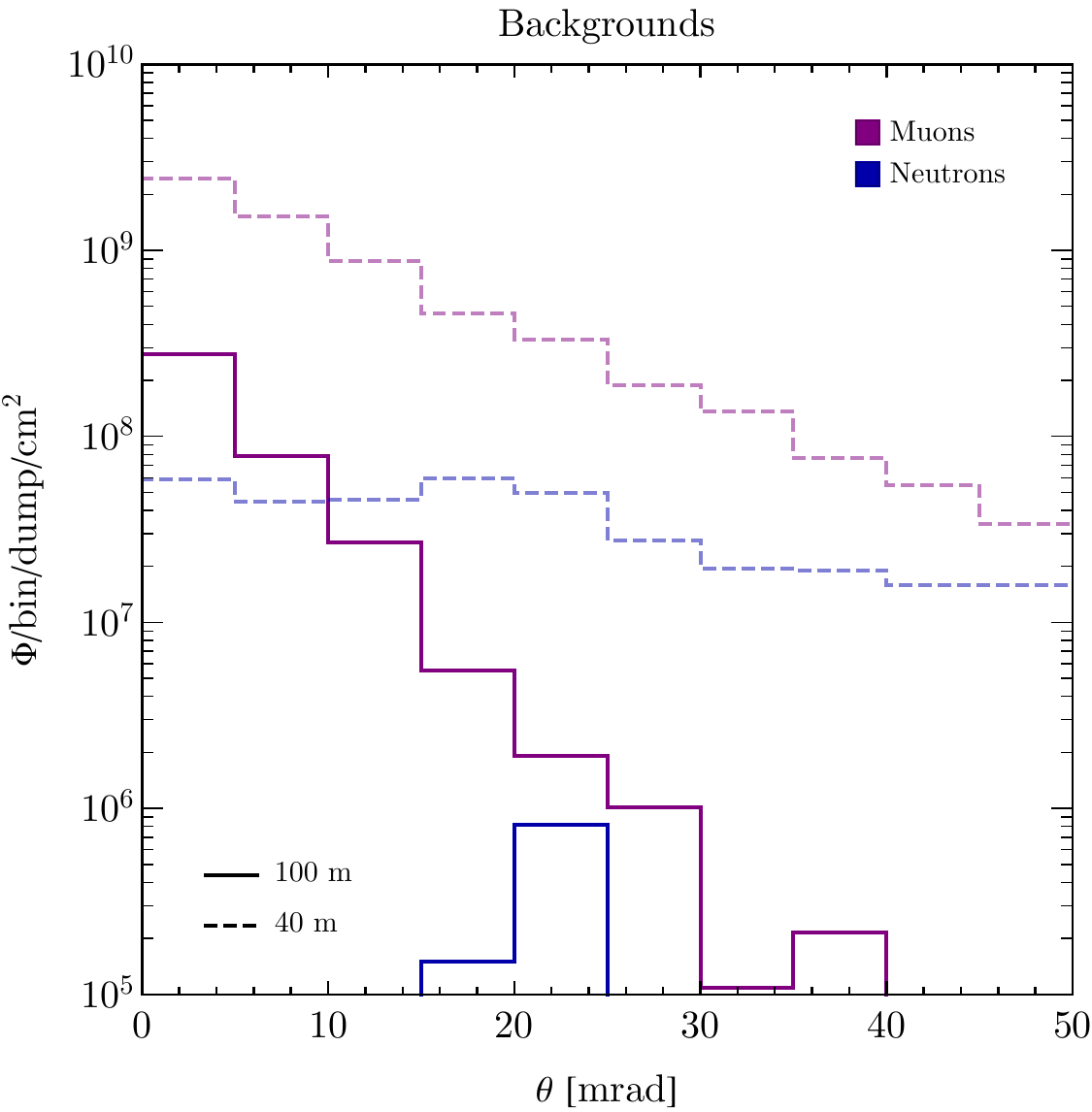}
\caption{Left: High-energy neutrino flux ($E_\nu > 10\ \GeV$) as function of the angle relative to the beam direction for detectors at a position of 100 m from the interaction point. Right: Forward muon (purple) and neutron (dark blue) flux as function of the angle relative to the beam direction. Full (dashed) lines indicate the flux at 100 (40) m from the interaction point.}
\label{fig:HE-theta}
\end{figure}
%%%%%%%%%%%%%%%%%%%%%%%%%%%%%%%%%%%%%%%%%

To better understand the forward component of the signal, we show in the left panel of  Fig.~\ref{fig:HE-theta} the angular dependence of the high energy ($10<E_\nu<100$~GeV) and very high energy ($E_\nu>100$~GeV) fluxes of neutrinos 100~meters away from the dump.
We add neutrinos and antineutrinos of the same flavor in these histograms.
We see that the flux is dominated by $\nu_\mu/\bar\nu_\mu$, as expected, with a small $\nu_\tau/\bar\nu_\tau$ component.
The drop in the high energy neutrino flux with off-axis angle is simply a matter of kinematics.
We also see that neutrinos above 100~GeV are mostly in the forward direction, within 5-10~mrad.
In the right panel of Fig.~\ref{fig:HE-theta}, we show the flux of muons and neutrons for distances of 40 meters (dashed) and 100 meters (solid) from the dump.
We highlight two important features in this plot.
First, while the muon flux is mostly forward, the neutron flux tends to be more isotropic, as the latter background is dominated by lower energy particles.
Second, even at 100~meters there is a considerable flux of high energy muons.

\begin{figure}[t]
\centering
\includegraphics[width=\linewidth]{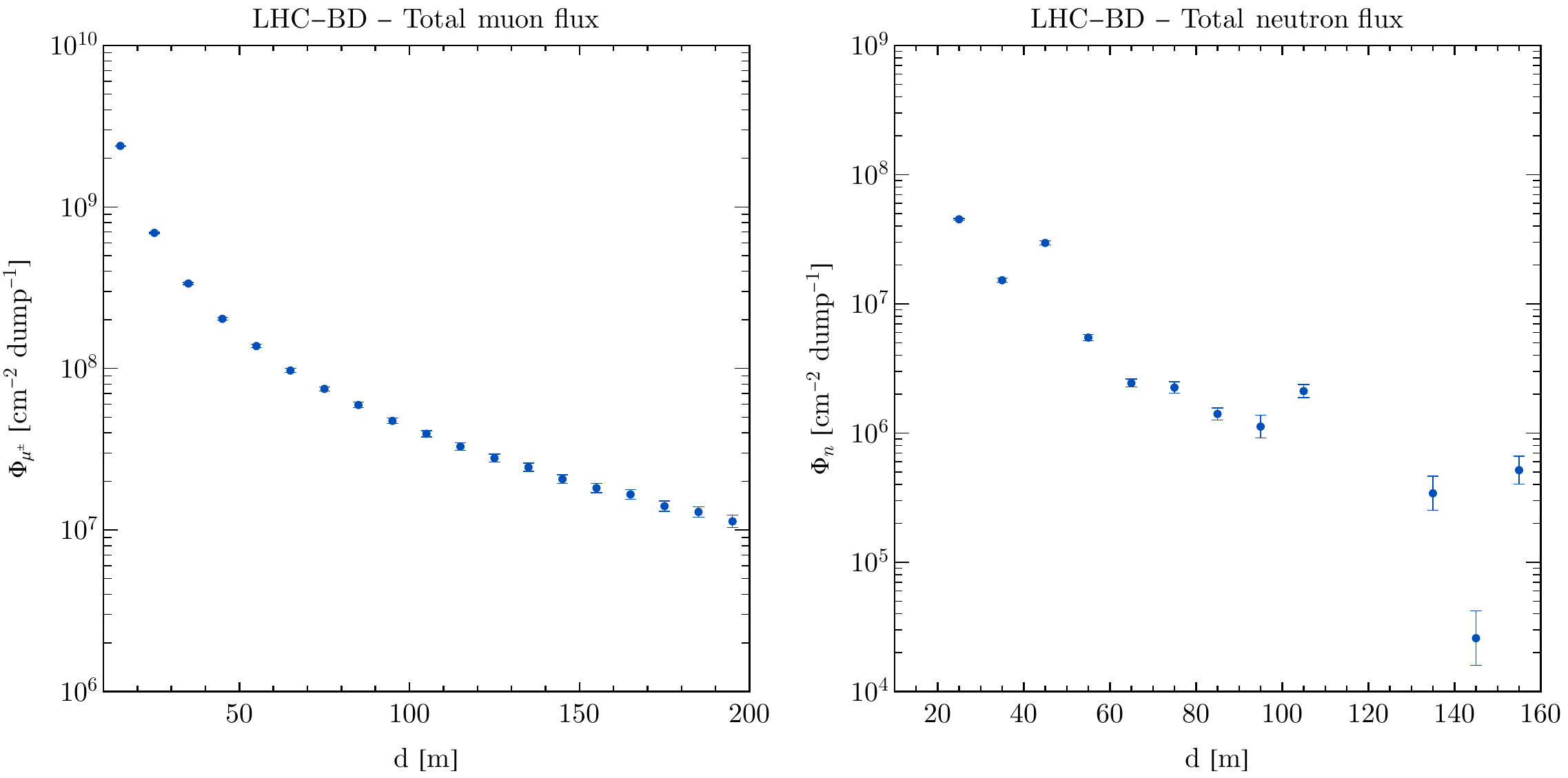}
\caption{Total muon (left) and neutron (right) fluxes per ${\rm cm}^2$ per beam dump as function of the distance to the main LHCBD cavern in the forward direction.  
\label{fig:MuonBkg}}
\end{figure}

\begin{figure}[t]
\centering
\includegraphics[width=0.95\linewidth]{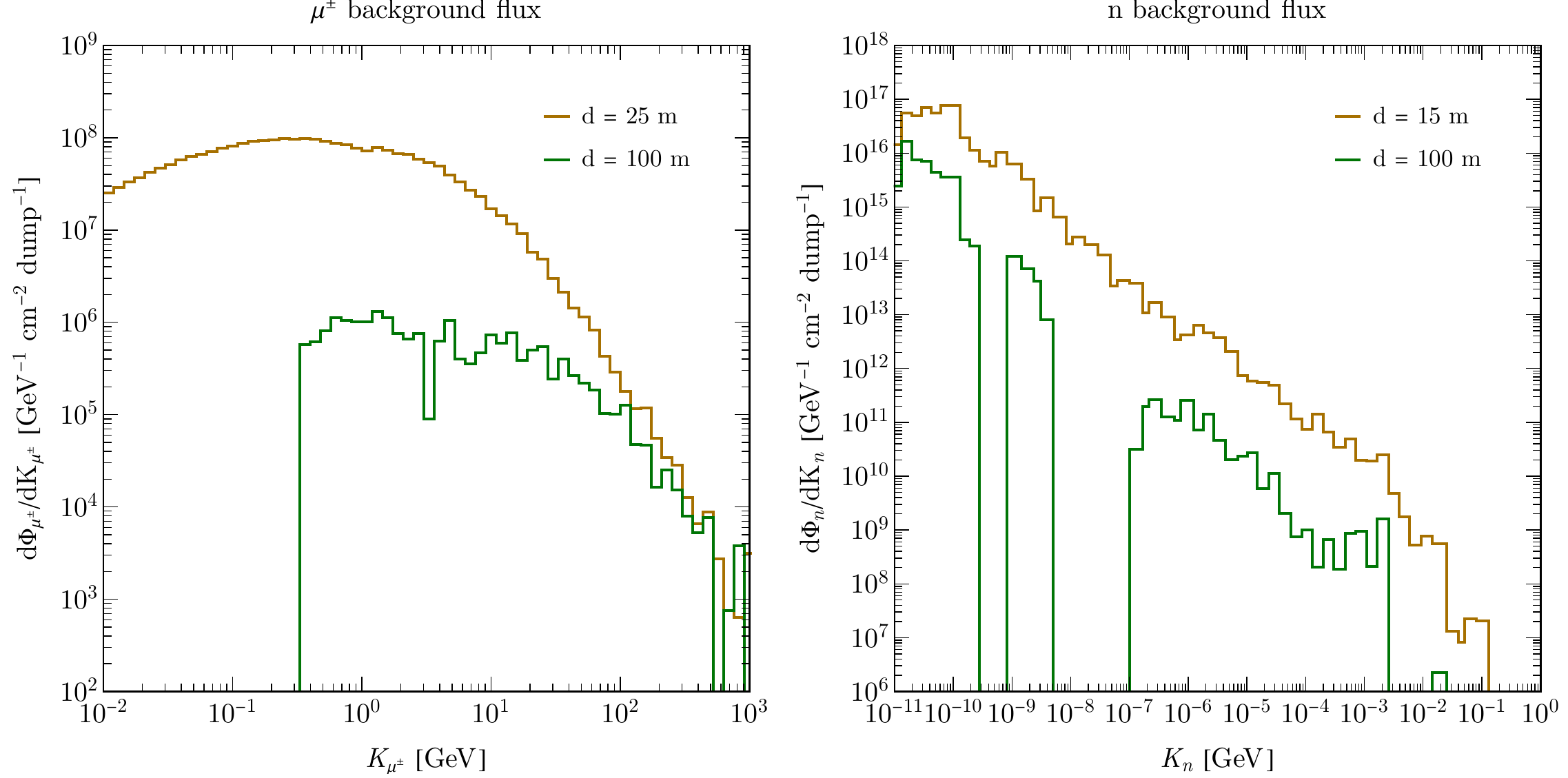}
\caption{Left: kinetic energy spectrum of muons at distances of $25$~m (orange) and $100$~m (green) from the beam dump.  
Right: kinetic energy spectrum of neutrons at distances of $15$~m (orange) and $100$~m (green) from the beam dump. 
\label{fig:BkgSpec}}
\end{figure}

To better understand the backgrounds in the forward direction, in  Fig.~\ref{fig:MuonBkg}, we present the total muon (left panel) and neutron (right panel) fluxes, on the beam axis direction, as function of the distance to the beam dump facility. 
The error bars are obtained from the {\tt FLUKA} simulation and consist of MC statistical uncertainties. 
The muon flux in the forward direction consists mostly of high energy particles.
We show the kinetic energy spectrum of both muon and neutron backgrounds in the forward direction for two different distances in Fig.~\ref{fig:BkgSpec}. 
In particular, the high energy component of the muon flux is not attenuated much by the soil in front of the cavern.
Dealing with this background would require an active shielding to deflect the muons.

\section{Coherent Scattering Measurements with Zero Background}
\label{app:CEvNSZeroBackground}
In Section~\ref{subsubsec:CohBackgrounds}, we discussed the expected background rate from neutron scattering events in the coherent scattering measurement we have proposed. We argued there that this relatively large (signal-to-background ratio of ${\sim}1$) background may be nearly completely mitigated with 3 m of additional shielding, placing the detector at $18$ m from the LHC Beam Dump instead of $15$ m. This results in the (isotropic) neutrino flux, and therefore the signal event rate, dropping by a factor of $(15/18)^2 \approx 70\%$, and a background-free search. In reality, there would still be some small prompt neutron flux, and in this scenario, the NIN contribution may actually be nonzero. We leave these details, as well as discussion of optimizing the detector location, to future dedicated studies.

\begin{figure}[th]
\begin{center}
\includegraphics[width=0.7\linewidth]{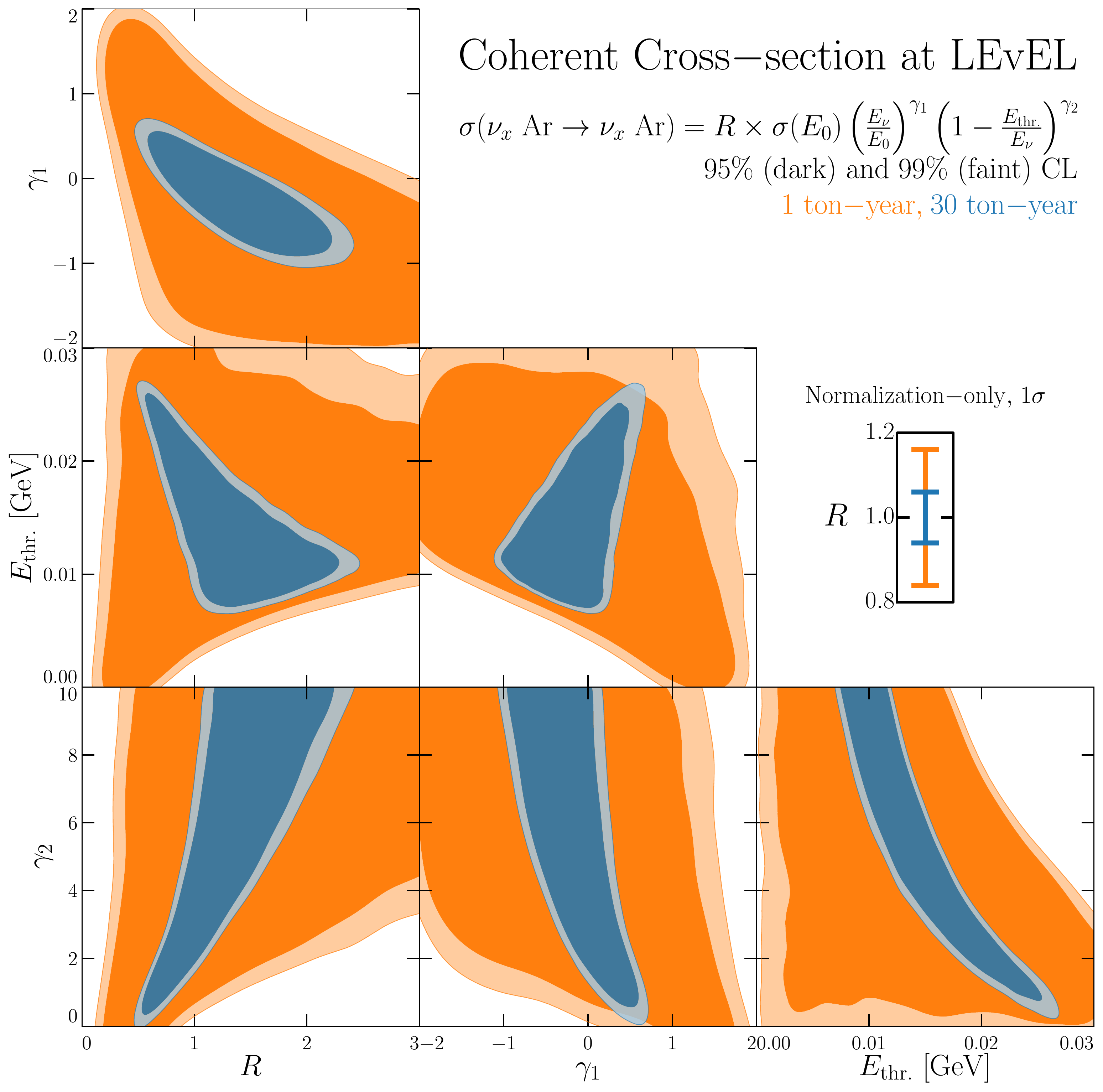}
\caption{Similar to Fig.~\ref{fig:NuCohXSec}, the measurement capability of LEvEL according to the parameterization of Eq.~\eqref{eq:cohXSec}, now with a zero-background assumption. The detector is assumed to be a distance of 18 m from the LHC Beam Dump instead of 15 m, resulting in a smaller signal rate. Systematic uncertainties ($5\%$ overall flux uncertainty and $10\%$ kaon-to-pion production ratio uncertainty) are still included and marginalized over.
\label{fig:NuCohXSec_NoBkg}}
\end{center}
\end{figure}

\begin{figure}[!ht]
\begin{center}
\includegraphics[width=0.8\linewidth]{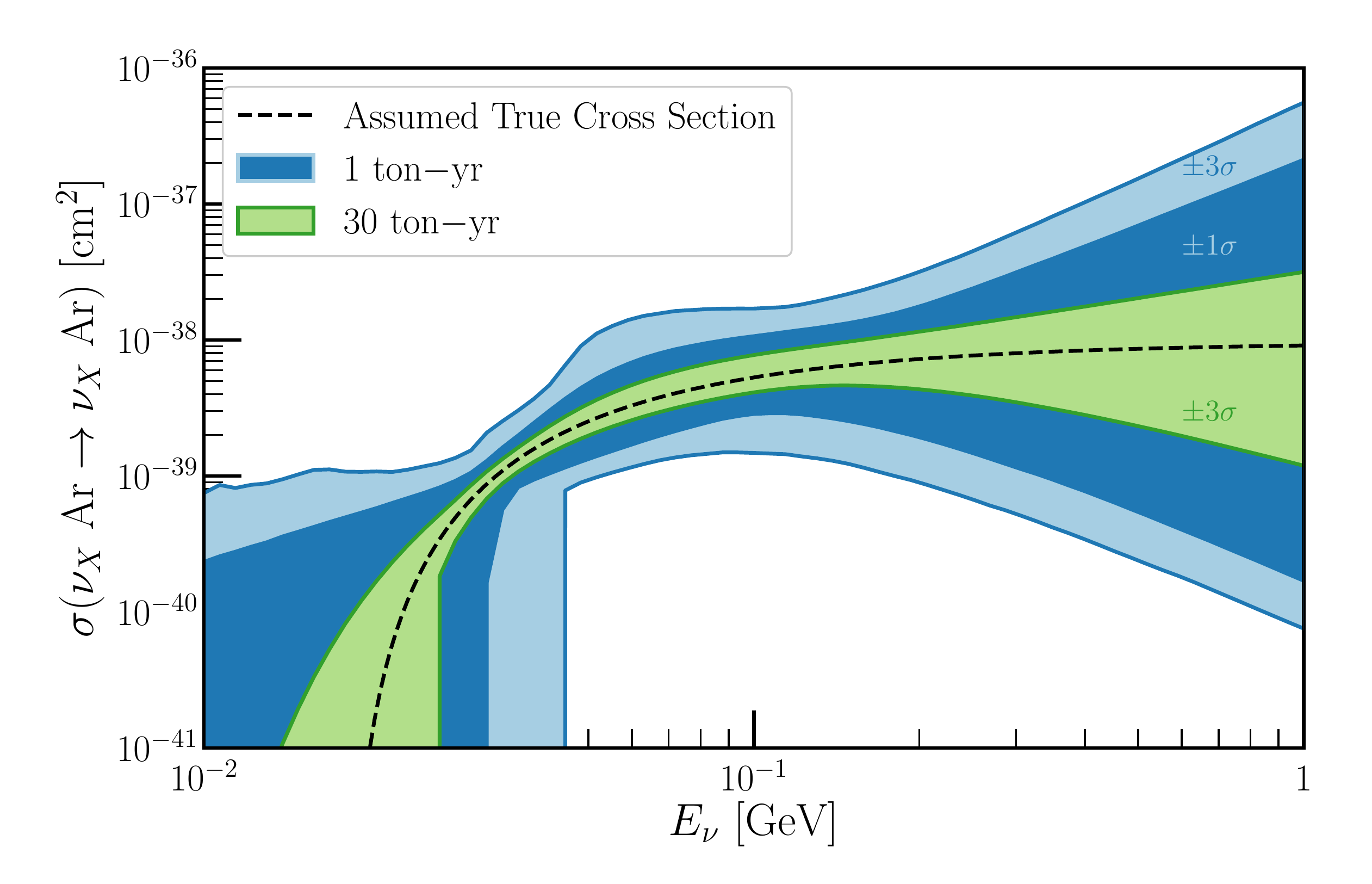}
\caption{Similar to Fig.~\ref{fig:CohXSecAllowed}, the implied measurement of the coherent cross section as a function of neutrino energy, now with a zero-background assumption. The detector is assumed to be a distance of 18 m from the LHC Beam Dump instead of 15 m, resulting in a smaller signal rate. Systematic uncertainties ($5\%$ overall flux uncertainty and $10\%$ kaon-to-pion production ratio uncertainty) are still included and marginalized over.
\label{fig:CohXSecAllowed_NoBkg}}
\end{center}
\end{figure}

Here, we present the same results as in Section~\ref{subsubsec:CohSystMeasurement} but under this alternate proposal. Fig.~\ref{fig:NuCohXSec_NoBkg} is identical to Fig.~\ref{fig:NuCohXSec} except now, the expected background is taken to be zero. The measurement capability of two exposures (1 ton-yr in orange and 30 ton-yr in blue) is shown for the different parameters of interest here. Similarly, Fig.~\ref{fig:CohXSecAllowed_NoBkg} repeats the analysis cf Fig.~\ref{fig:CohXSecAllowed} in the main text under this alternate-background assumption. The measurement capability of the cross section as a function of $E_\nu$ here is moderately better than in Fig.~\ref{fig:CohXSecAllowed}, although the improvement is not too dramatic.

%%%%%%%%%%%%%%%%%%%%%%%%%%%%%%%%%%%%%%%%%%%%%%%%%%%%%%%%%%%%%%%%%%%%%%%%%%%%%%%%%%
\bibliographystyle{JHEP}
\bibliography{refs}

\end{document}